\documentclass[12pt]{article}
\usepackage[top=2cm,bottom=3cm,left=2cm,right=2cm]{geometry}
\usepackage{amsmath}
\usepackage{amssymb} 
\usepackage{graphicx}
\usepackage{hyperref}
\usepackage[mathscr]{euscript}

\numberwithin{equation}{section}
\numberwithin{figure}{section}

\def\eq#1{(\ref{eq:#1})}
\def\lineup{\!\!\!\!\!\!\!\! &&}

\newcommand{\Tr}{\mathop{\rm Tr}\nolimits}

\def\d{\partial}
\def\eps{\epsilon}

\def\sgn{\mathrm{sgn}}

\begin{document}
\begin{titlepage}
\rightline\today

\begin{center}
\vskip 3.5cm

{\large \bf{Rolling Near the Tachyon Vacuum}}

\vskip 1.0cm

{\large {Theodore Erler\footnote{tchovi@gmail.com}, Toru Masuda\footnote{masudatoru@gmail.com}, Martin Schnabl\footnote{schnabl.martin@gmail.com}}}

\vskip 1.0cm

{\it Institute of Physics of the ASCR, v.v.i.}\\
{\it Na Slovance 2, 182 21 Prague 8, Czech Republic}\\




\vskip 2.0cm

{\bf Abstract} 

\end{center}

In a linear dilaton background, it has been argued that an unstable D-brane can decay to the tachyon vacuum without leaving behind a remnant of tachyon matter. Here we address the question of how the D-brane can decay to the tachyon vacuum when the tachyon vacuum does not support physical fluctuations. Using the formalism of open string field theory, we find that the tachyon vacuum {\it can} support fluctuations provided they are ``hidden" as nonperturbative effects behind a pure gauge asymptotic series.

\end{titlepage}

\tableofcontents
 
\section{Motivation}

String field theory provides an interesting framework for understanding the nature of time and vacuum selection in string theory. Particularly important in this respect are time dependent classical solutions---in open bosonic string field theory, especially solutions describing the decay of unstable D-branes. In the string field theory literature, the most widely studied example is the time-like, homogeneous decay process \cite{MoellerZwiebach,Hata,Schnabl_marg,KORZ,Ellwood,KOS,Longton}, corresponding to the rolling tachyon boundary deformation of Sen \cite{SenRolling}.

However, perhaps the simplest time-dependent solution is the light-like rolling tachyon deformation described by Hellerman and Schnabl~\cite{Hellerman}, following sigma model analysis of analogous deformations of the closed string tachyon \cite{Hellerman1,Hellerman2,Hellerman3}. This solution assumes a D-brane whose worldvolume contains a copy of 2D Minkowski space $(x^0,x^1)$ carrying a light-like linear dilaton $\Phi_\mathrm{dilaton} = -\frac{1}{\beta} x^-$.\footnote{We assume $\alpha'=1$ and mostly plus metric signature. The dilaton gradient need not be light-like if we are interested in noncritical strings.} In this background, the vertex operator $e^{\beta X^+}(z)$ is exactly marginal, and generates an open string field theory solution where, in lightcone time $x^+$, the D-brane decays starting from an infinitesimal, homogeneous tachyon fluctuation in the infinite past. The remarkable fact, argued in \cite{Hellerman}, is that towards the infinite future the solution approaches a state where the D-brane has disappeared, leaving only empty space---the ``tachyon vacuum." This is quite different from the usual time-like rolling tachyon deformation, where the final state---dubbed ``tachyon matter"~\cite{SenTM}---does not appear to represent a finite and stationary configuration from the perspective of open string degrees of freedom. 

However, the light-like deformation raises a paradox. The tachyon vacuum does not allow for small fluctuations, since there are no D-branes to support open string states. But this seems to imply that it is impossible to reach the tachyon vacuum by a continuous physical process. Let us frame this issue more specifically in the context of open string field theory. The decay of the D-brane can be described by an exact solution of Witten's open bosonic string field theory in $\mathcal{B}_0$ gauge \cite{Schnabl_marg,KORZ}:
\begin{equation}
\Psi = \sqrt{\Omega}c e^{\beta X^+} \frac{B}{1+\frac{1-\Omega}{K}e^{\beta X^+}}c\sqrt{\Omega}, \label{eq:Schgauge}
\end{equation}
where the string field $e^{\beta X^+}$ represents an insertion of the marginal operator $e^{\beta X^+}(z)$ in correlation functions on the cylinder.\footnote{We use the algebraic formalism introduced in \cite{Okawa} and using the conventions of \cite{simple}. The objects $K,B,c,e^{\beta X^+}$ are string fields multiplied with Witten's open string star product, and $\Omega=e^{-K}=|0\rangle$ is the $SL(2,\mathbb{R})$ vacuum.} The conventional interpretation of the solution is as an expansion in powers of~$e^{\beta X^+}$:
\begin{eqnarray}
\Psi \lineup = \sqrt{\Omega}\, ce^{\beta X^+}\sqrt{\Omega}-\sqrt{\Omega}\, cB e^{\beta X^+}\frac{1-\Omega}{K}\, e^{\beta X^+} c\sqrt{\Omega}\nonumber\\
\lineup\ \ \ \ \ \ \ \ \ \ \ \ \ \ \ \ \ \ \ \ \ \ +\sqrt{\Omega} \,cB e^{\beta X^+}\frac{1-\Omega}{K}\,e^{\beta X^+}\frac{1-\Omega}{K} e^{\beta X^+}\,c \sqrt{\Omega}-...\ .
\label{eq:initial}\end{eqnarray}
This is a good approximation at early times, when $e^{\beta X^+}$ is small. For $x^+\to -\infty$, the solution vanishes, representing the original unstable D-brane configuration. However, since $X^+(z)$ has trivial self-contractions, we can also expand the solution using inverse powers of~$e^{\beta X^+}$:
\begin{eqnarray}
\Psi \lineup = \sqrt{\Omega}c\frac{KB}{1-\Omega}c\sqrt{\Omega}-\sqrt{\Omega}cB\frac{K}{1-\Omega}e^{-\beta X^+}\frac{K}{1-\Omega}c\sqrt{\Omega}\nonumber\\
\lineup \ \ \ \ \ \ \ \ \ \ \ \ \ \ \ \ \ \ \ \ \ \ \ \ +\sqrt{\Omega}cB\frac{K}{1-\Omega}e^{-\beta X^+}\frac{K}{1-\Omega}e^{-\beta X^+}\frac{K}{1-\Omega}c\sqrt{\Omega}-...\ .
\label{eq:final}\end{eqnarray}
This should be a good approximation at late times, when $e^{-\beta X^+}$ is small. For $x^+\to\infty$, only the leading term survives, and is precisely the same as the $\mathcal{B}_0$ gauge solution for the tachyon vacuum~\cite{Schnabl}. The subleading terms appear to represent fluctuations of the tachyon vacuum. However, they are not genuine fluctuations; they can be removed by gauge transformation, order by order in powers of $e^{-\beta x^+}$. This seems to imply that the time dependence of the solution---at least as expressed in \eq{final}---is a gauge artifact. One way this could be reconciled is if the decay process terminates at the tachyon vacuum at a finite moment, exactly when the expansion \eq{final} begins to converge. Thereafter we would only have empty space, free of disturbances from D-branes and open strings. However, this would imply that no measurement around the tachyon vacuum could determine if D-brane decay had occurred in the past. This would be a clear demonstration that string theory admits highly acausal dynamics. Despite the appearance of infinite order time derivatives in the string field theory action, this seems to contradict our physical understanding of string theory in known backgrounds. 

The above difficulty is related to the apparent impossibility of getting ``close" to the tachyon vacuum when the tachyon vacuum does not admit physical deformation. The goal of this paper is to understand the resolution of this paradox. Our results can be summarized as follows:
\begin{itemize}
\item Using the analytic solution of Kiermaier, Okawa, and Soler \cite{KOS}, we are able to determine the exact evolution of the string field in lightcone time. We find that as $x^+\to-\infty$ the solution vanishes, representing the original unstable D-brane configuration, and as $x^+\to+\infty$ the solution becomes identical to the tachyon vacuum solution of \cite{simple}. Previous computations of the solution have relied on level truncation and the perturbative expansion in powers of~$e^{\beta x^+}$ \cite{Hellerman,Moeller}. Probing the late time behavior in this way requires very high order computations,\footnote{Hellerman and Schnabl computed the tachyon profile in level $0$ truncated open string field theory including perturbative corrections up to order $(e^{\beta x^+})^{5000}$.} and while the results reasonably confirm expectations, they are not fully conclusive. 
\item We find that the solution can be approximated by a perturbative expansion around the tachyon vacuum in powers of $e^{-\beta x^+}$ at late times. However---and this is the central point---the expansion is asymptotic, with vanishing radius of convergence in powers of $e^{-\beta x^+}$. Therefore, even though the corrections in the late time expansion are order-by-order pure gauge, it does not necessarily follow that the solution is gauge equivalent to the tachyon vacuum at any finite time. 
\item The asymptotic expansion is not Borel resummable. This indicates the presence of nonperturbative corrections to the decay process at late times. We argue that the corrections represent a genuine ``physical" fluctuation of the tachyon vacuum which persists through the decay process and finally vanishes in the infinite future. In particular, the light-like rolling tachyon background is not equivalent to the tachyon vacuum at any finite time.
\end{itemize}
In essence, we find that the tachyon vacuum {\it does} support small fluctuations provided they are hidden as nonperturbative effects behind a pure gauge asymptotic series.

This paper is organized as follows. In section \ref{sec:profile} we describe the solution and compute the trajectory of the tachyon field as a function of lightcone time $x^+$. We confirm that the solution does approach the tachyon vacuum in the distant future. In section \ref{sec:final} we consider the late time expansion around the tachyon vacuum. To simplify computation we analyze a factor of the solution at ghost number $0$ which carries all of the dependence on lightcone time. We show that the analogue of the tachyon at ghost number $0$ can be expanded at late times as a power series in $e^{-\beta x^+}$ with vanishing radius of convergence. We also find that the expansion is not Borel resummable. In section \ref{sec:wrong} we investigate what happens when the D-brane decays in the ``wrong direction," towards the side of the tachyon effective potential which is unbounded from below. We find that the ghost number $0$ tachyon diverges super-exponentially at late times, as determined by a sliver-like critical point in the integration over wedge states. This behavior contradicts a naive expectation that the late time behavior should be described by a perturbative expansion around the tachyon vacuum, but with alternating signs relative to the case of section \ref{sec:final}. In fact, the expansion with alternating signs is Borel resummable, and in this way we can reconstruct a surprising solution which appears to decay to the tachyon vacuum from the opposite side of the local minimum. We find that the solution is characterized by wedge states with negative width, and is therefore not normalizable. In section \ref{sec:Stokes}, we use steepest descent analysis to show that the sliver-like critical point analyzed in the previous section is in fact responsible for a super-exponentially small nonperturbative contribution to the late time behavior when the string field approaches the tachyon vacuum. In section \ref{sec:physical} we argue that this nonperturbative contribution represents a physical fluctuation of the tachyon vacuum. To make this claim we must analyze a gauge invariant quantity; we consider the amplitude for emission of an on-shell closed string tachyon from the decaying D-brane. We find that the asymptotic expansion around the tachyon vacuum makes vanishing contribution to this amplitude, as would be expected since the expansion is order-by-order pure gauge. To talk meaningfully about nonperturbative effects, however, the perturbative expansion must be present in some form. For this reason we consider a controlled deformation of the amplitude which breaks gauge invariance. In the deformed amplitude we can identify a contribution from the late time asymptotic expansion and a nonperturbative contribution from a sliver-like critical point. Taking the deformation parameter to zero, we find that the nonperturbative contribution accounts for the entire nonvanishing amplitude. We further apply this analysis to the solution of section \ref{sec:wrong} which rolls to the tachyon vacuum from the opposite side. Since the solution is defined entirely through Borel summation, nonperturbative effects are not present, and the amplitude therefore vanishes when the deformation parameter is taken to zero. This indicates that the solution is a time-dependent gauge transformation of the tachyon vacuum. Interestingly, this confirms the intuition derived from the tachyon effective potential of boundary string field theory, where the local minimum sits at infinity and it is not meaningful to ask what lies beyond. In section \ref{sec:resurgence} we show that the late time expansion around the tachyon vacuum is resummable in the more general sense of resurgence theory once we account for contributions from nonperturbative saddle points. In section \ref{sec:gauge} we investigate a formal gauge transformation which naively maps the light-like rolling tachyon solution into the tachyon vacuum, and explain why the proposed gauge transformation breaks down. We end with discussion of future directions. 
 
\section{Tachyon Profile}
\label{sec:profile}

The light-like rolling tachyon deformation assumes a D-brane whose worldvolume contains a copy of 2D Minkowski space $(x^0,x^1)$ carrying a light-like linear dilaton
\begin{equation}\Phi_\mathrm{dilaton} = -\frac{1}{\beta}x^-,\end{equation}
where $\beta>0$ is a parameter and lightcone coordinates are defined
\begin{equation}x^\pm = \frac{1}{\sqrt{2}}(x^0\pm x^1).\end{equation}
The open string worldsheet theory contains a factor given by a noncompact, time-like free boson $X^0(z,\overline{z})$ together with noncompact, space-like free boson $X^1(z,\overline{z})$ both subject to Neumann boundary conditions and in the presence of the light-like linear dilaton. The elementary correlator is the 1-point function of a boundary plane wave vertex operator in the upper half plane:
\begin{equation}
\big\langle e^{i k\cdot X}(z)\big\rangle_\mathrm{UHP}^{X^0,X^1}=(2\pi)^2\delta^2(k+i V),
\end{equation}
where $V_\mu$ is the dilaton gradient 
\begin{equation}V_\mu = -\frac{1}{\beta}\delta^-_\mu,\end{equation}
and we introduce basis 1-forms $\delta^\pm_\mu$ defined to satisfy $\delta^\pm \cdot x =\delta_\mu^\pm x^\mu = x^\pm$ for any vector $x^\mu$. The boundary plane wave vertex operator $e^{i k\cdot X}(z)$ has conformal dimension $k^2+i V\cdot k$. In particular, $e^{\beta X^+}(z)$ has conformal dimension $1$ and is a marginal operator, while $e^{-\beta X^+}(z)$ has conformal dimension $-1$. The worldsheet theory contains two additional factors: the usual $bc$ ghost system of central charge $-26$, and a factor in the matter boundary conformal field theory of central charge $24$. Since the light-like rolling tachyon background does not excite primaries in the $c=24$ component, its precise form is not important for our considerations.\footnote{We normalize the ghost correlator as 
\begin{equation}\Big\langle c\d c\d^2 c(0)\Big\rangle_\mathrm{UHP}^{bc}=-2,\end{equation} and in the $c=24$ factor we assume $\langle 1\rangle_{\mathrm{UHP}}^{c=24}=1$.}

The light-like rolling tachyon background is defined by perturbing the boundary of the worldsheet by the exactly marginal operator
\begin{equation}\lambda e^{\beta X^+}(0).\end{equation}
This operator is exactly marginal since it has trivial self-contractions. The constant $\lambda$ parameterizes the strength of the boundary interaction; without loss of generality we can assume $|\lambda|=1$, since any other value can be absorbed into a redefinition of the origin of the lightcone coordinate $x^+$. However the sign of $\lambda$ is important, since it determines whether the background rolls towards the tachyon vacuum ($\lambda>0$) or whether it rolls away from the tachyon vacuum, in the direction where the tachyon effective potential is unbounded from below ($\lambda<0$). Our primary interest is rolling to the tachyon vacuum, but the solution for $\lambda<0$ is also important and will be discussed in section~\ref{sec:wrong}.

Our analysis is based on the analytic solution of Kiermaier, Okawa, and Soler \cite{KOS}, which can be written in the form \cite{KOSsing}:
\begin{equation}
\Psi = c(1+K)B c\frac{1}{1+K} - c(1+K)\sigma\frac{B}{1+K}\overline{\sigma}(1+K)c\frac{1}{1+K}.\label{eq:KOS}
\end{equation}
The string fields $\sigma,\overline{\sigma}$ represent boundary condition changing operators which shift from the original unstable D-brane to the light-like rolling tachyon background. The solution is physically equivalent to the $\mathcal{B}_0$ gauge solution studied in \cite{Hellerman}, but is more convenient for computations. Note that the solution is a sum of two terms, and the first term is by itself a solution for the tachyon vacuum~\cite{simple}. This means that the decay process will approach the tachyon vacuum at late times if the boundary condition changing operators $\sigma,\overline{\sigma}$ vanish in the infinite future. This is actually natural. Boundary condition changing operators represent stretched strings between D-branes; in the current situation, they represent strings connecting the original unstable D-brane to a copy of itself in the process of decay. If the final product of decay is the tachyon vacuum, the stretched strings should disappear, and the boundary condition changing operators should vanish. In the infinite past, the light-like rolling tachyon background is indistinguishable from the original D-brane, and we expect that $\sigma=\overline{\sigma}=1$. Then the two terms in \eq{KOS} cancel, giving the trivial solution $\Psi=0$.

The most concrete way to characterize the time evolution of the string field is through the evolution of the infinite tower of ordinary fields which appear as coefficients of the expansion into a basis of $L_0$ eigenstates. The field with the lowest mass-squared is the tachyon:
\begin{equation}
\Psi = \int \frac{d^2 k}{(2\pi)^2}T(k)\, ce^{ik\cdot X}(0)|0\rangle + ...,
\end{equation}
where $T(k)$ is the tachyon field in momentum space and the dots indicate states with higher $L_0$ eigenvalue for a given momentum. In the following we focus on the tachyon, since the analysis of higher coefficients is similar. In position space, we expect that the tachyon field $T(x)$ will only be a function of lightcone time $x^+$. Moreover, if the solution approaches the tachyon vacuum, the tachyon field will be a smooth function of $x^+$ satisfying the boundary conditions
\begin{eqnarray}
\lim_{x^+\to -\infty} T(x) \lineup = 0,\\
\lim_{x^+\to+\infty} T(x) \lineup = .2844...,
\end{eqnarray} 
where $.2844...$ is the expectation value of the tachyon field at the tachyon vacuum, as defined by the solution of \cite{simple}. 

The second term in the solution \eq{KOS} can be expressed in terms of wedge states:
\begin{equation} \left(1-\left.\frac{\d}{\d \eps_1}\right|_{\eps_1=0}\right)\left(1-\left.\frac{\d}{\d \eps_2}\right|_{\eps_2=0}\right)\int_0^\infty dt_1\int_0^\infty dt_2\, e^{-t_1-t_2}\, c B\,\Omega^{\eps_1}\,\sigma\, \Omega^{t_1}\,\overline{\sigma}\,\Omega^{\eps_1}\,c\,\Omega^{t_2},
\end{equation}
where $t_1,t_2$ arise as Schwinger parameters in the formula
\begin{equation}\frac{1}{1+K} = \int_0^\infty dt\, e^{-t}\Omega^t.\label{eq:Schwinger}\end{equation}
Using standard manipulations, this leads to the following expression for the tachyon field:
\begin{eqnarray}
T(x) \lineup = .2844...+\frac{\pi}{2}\left(1-\left.\frac{\d}{\d \eps_1}\right|_{\eps_1=0}\right)\left(1-\left.\frac{\d}{\d \eps_2}\right|_{\eps_2=0}\right)\int_0^\infty dt_1\int_0^\infty dt_2 e^{-t_1-t_2}\nonumber\\
\lineup\ \ \ \ \ \ \ \ \ \ \ \ \ \ \ \  \times \int\frac{d^2 k}{(2\pi)^2} e^{ik\cdot (x- V \ln \frac{\pi}{2})}\Tr\Big[\Omega^{\frac{1}{2}+\eps_1}\,\sigma\, \Omega^{t_1}\,\overline{\sigma}\,\Omega^{\frac{1}{2}+\eps_2+t_2} e^{-i(k+iV)\cdot X}\Big]^{X^0,X^1}\nonumber\\
\lineup\ \ \ \ \ \ \ \ \ \ \ \ \ \ \ \  \times \Tr\Big[\Omega^{\frac{1}{2}}\,c B\,\Omega^{\eps_1+\eps_2+t_1}\, c\,\Omega^{\frac{1}{2}+t_2}\,c \d c\Big]^{bc},\label{eq:T}
\end{eqnarray}
where the traces represent correlation functions, factorized into matter and ghost contributions. The ghost correlator is well known \cite{Okawa,Schnabl}:
\begin{equation}
\Tr\Big[\Omega^{\alpha_1}\,c B\,\Omega^{\alpha_2} \,c\,\Omega^{\alpha_3}\, c \d c\Big]^{bc} = -\frac{L^2}{\pi^3}\Big[\theta_{\alpha_1}\sin^2\theta_{\alpha_3}+\theta_{\alpha_3}\sin^2\theta_{\alpha_1}-\sin\theta_{\alpha_1}\sin\theta_{\alpha_2}\sin\theta_{\alpha_3}\Big],\label{eq:gh}
\end{equation}
where on the right hand side we introduce the variables 
\begin{eqnarray}
L \lineup = \alpha_1+\alpha_2+\alpha_3\\
\theta_{\alpha_i}\lineup =\frac{\pi \alpha_i}{L}.\label{variables}
\end{eqnarray}
The matter correlator is the important ingredient. In appendix \ref{app:bcc} we show that
\begin{equation}
 \int\frac{d^2 k}{(2\pi)^2} e^{ik\cdot (x- V \ln \frac{\pi}{2})}\Tr\Big[\Omega^{\alpha_1}\,\sigma\, \Omega^{\alpha_2}\,\overline{\sigma}\,\Omega^{\alpha_3} e^{-i(k+iV)\cdot X}\Big]^{X^0,X^1}= \exp\left[-e^{\beta x^+} \frac{2\lambda}{L}\frac{\sin \theta_{\alpha_2}}{\sin\theta_{\alpha_1}\sin\theta_{\alpha_3}}\right].\label{eq:bcc}
\end{equation}
Therefore we have a closed form expression for $T(x)$ which can be evaluated numerically for any value of $x^+$. This is an advantage over the $\mathcal{B}_0$ gauge solution, where there is no efficient procedure for computing the tachyon profile, and the late time behavior can only be discussed using indirect and formal arguments. Note that the dependence of the tachyon field on $x$ appears exclusively through the $e^{\beta x^+}$ factor in the exponential on the right hand side of \eq{bcc}. Moreover, if $\lambda>0$ the argument of the exponential is negative. This confirms that,  in the infinite future, the boundary condition changing operators effectively vanish, and the solution approaches the tachyon vacuum. 

\begin{figure}
\begin{center}
\resizebox{3.4in}{2in}{\includegraphics{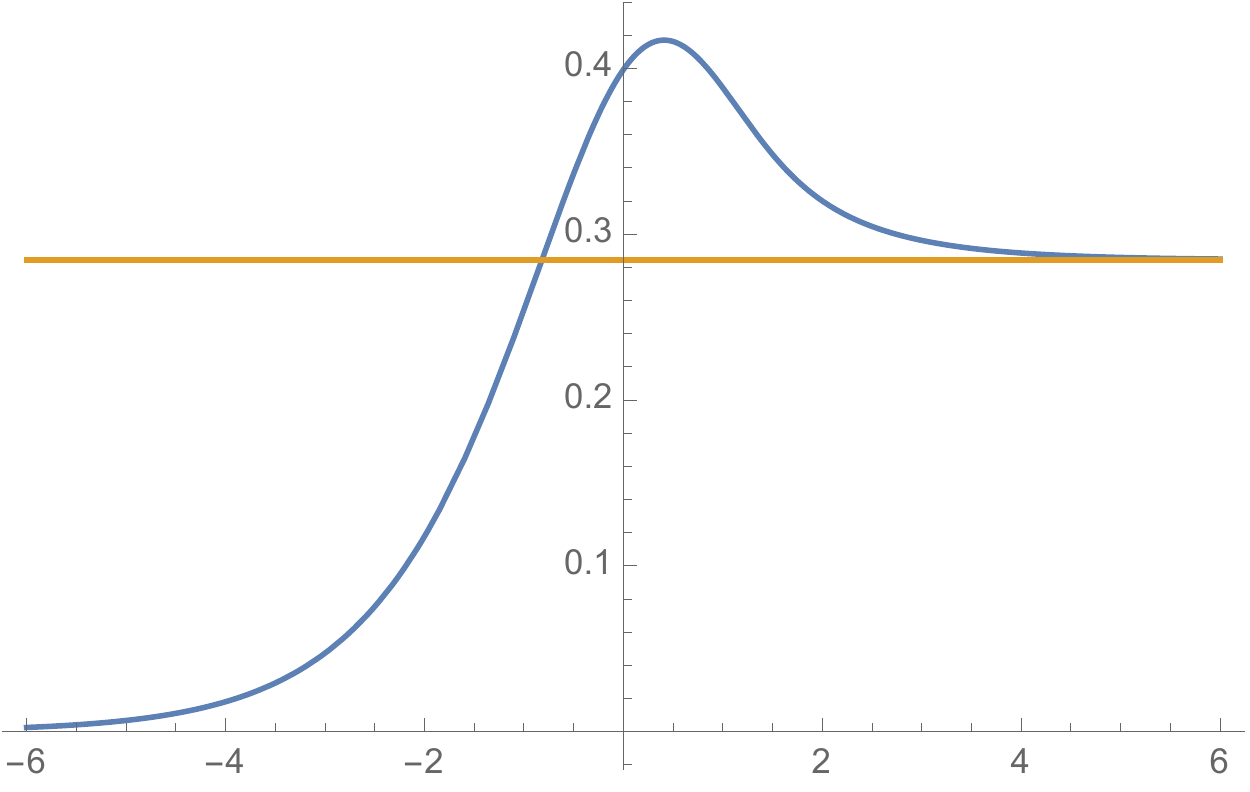}}
\end{center}
\caption{\label{fig:T} Profile of the tachyon field $T(x)$ as a function of $x^+\in[-6,6]$ for the light-like rolling tachyon solution. In this and subsequent plots we set $\beta=|\lambda|=1$. The flat line indicates the expectation value at the tachyon vacuum.}
\end{figure}

The evolution of the tachyon field is shown in figure \ref{fig:T}. As expected, there is a smooth transition from the unstable D-brane to the tachyon vacuum. One noteworthy feature is that the tachyon field overshoots, but then approaches the tachyon vacuum monotonically.\footnote{We also computed the tachyon profile for the real form of the solution, related to \eq{KOS} by similarity transformation of $\sqrt{1+K}$. Extracting the tachyon coefficient from this solution requires numerical evaluation of a more complicated three dimensional integral. The resulting profile is similar to \ref{fig:T}, but the tachyon does not overshoot the tachyon vacuum quite as dramatically.} This contrasts with level 0 computations \cite{Hellerman,Moeller}, where the tachyon undergoes damped oscillations around the tachyon vacuum at late times. The level 0 results in particular indicate complex exponential decay to the tachyon vacuum, whereas the exact solution produces real exponential decay. The computations of \cite{Moeller} however suggest that the oscillations are somewhat suppressed by the inclusion of higher level fields, so it is possible that figure \ref{fig:T} is more representative of the exact tachyon profile in Siegel gauge. 

\section{The Late Time Expansion}
\label{sec:final}

\begin{figure}
\begin{center}
\resizebox{3.4in}{2in}{\includegraphics{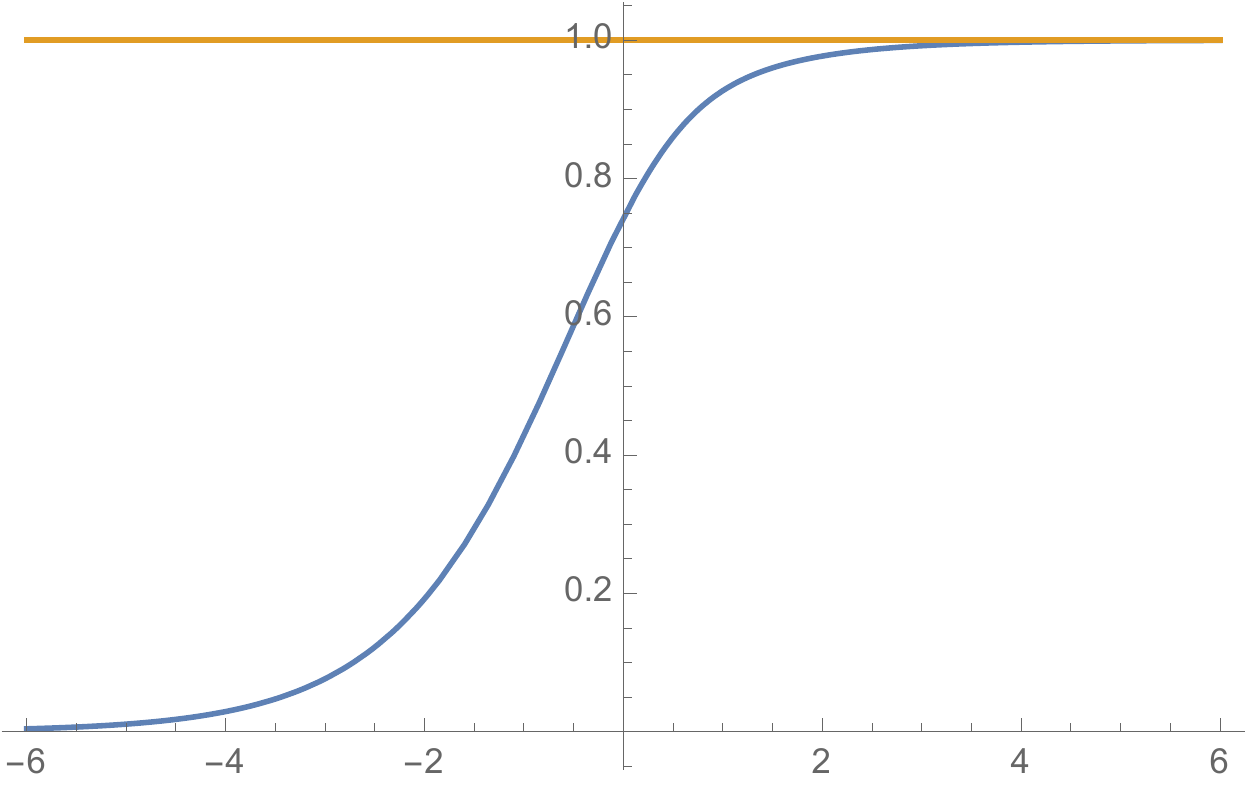}}
\end{center}
\caption{\label{fig:gamma} Profile of the ghost number zero tachyon  as a function of $x^+\in[-6,6]$. The flat line indicates the value $1$, which is the ghost number zero analogue of the expectation value at the tachyon vacuum.}
\end{figure}

We now want to understand how the solution approaches the tachyon vacuum. The solution can be written in the form
\begin{equation}
\Psi = cB (1+K)\left[\frac{1}{1+K}-\sigma \frac{1}{1+K}\overline{\sigma}\right](1+K)c\frac{1}{1+K}.
\end{equation}
Note that the dependence on time is completely contained in the factor in square brackets:
\begin{equation}\Gamma \equiv\frac{1}{1+K}-\sigma\frac{1}{1+K}\overline{\sigma}.\end{equation}
To give more palatable formulas we henceforth focus our analysis on the state $\Gamma$, rather than the full solution $\Psi$. The extra ghost insertions and wedge state factors in $\Psi$ do not alter the conclusions. At ghost number $0$, the state with lowest $L_0$ eigenvalue for a given momentum is 
\begin{equation}
\Gamma = \int \frac{d^2 k}{(2\pi)^2}\gamma(k) e^{ik\cdot X}(0)|0\rangle +...\ .
\end{equation}
The field $\gamma(x)$ can be viewed as a ghost number 0 analogue of the tachyon. Following standard manipulations and using \eq{bcc}, we find that $\gamma(x)$ evolves through the decay process according to 
\begin{equation}\gamma(x) = 1-\int_0^\infty dt \,e^{-t}\exp\left[-\frac{8\lambda}{\pi}e^{\beta x^+}\tau(t)\right],\label{eq:gammaint}\end{equation}
where 
\begin{equation}\tau(t)\equiv\frac{\pi}{2(t+1)}\cot \frac{\pi}{2(t+1)}.\label{eq:tau}\end{equation}
The function $\tau(t)$ vanishes at $t=0$ and increases monotonically towards $1$ as $t\to\infty$. The integration variable $t$ is the Schwinger parameter appearing in \eq{Schwinger}. In the infinite past, the integral cancels against the first term so that $\gamma(x)$ vanishes. This represents the original unstable D-brane. In the infinite future, the integral approaches zero assuming $\lambda$ is positive, so that $\gamma(x)$ approaches $1$. This represents the tachyon vacuum. We plot the profile of $\gamma(x)$ in figure \ref{fig:gamma}. 

To derive early and late time expansions, we represent the boundary condition changing operators $\sigma,\overline{\sigma}$ through integration of the marginal operator $e^{\beta X^+}$ on the boundary of the worldsheet. In this way, the state $\Gamma$ is written as \cite{KOS}:
\begin{equation}\Gamma = \frac{1}{1+K}-\frac{1}{1+K+\lambda e^{\beta X^+}}\label{eq:Gammaexp}\end{equation}
which leads to the early and late time expansions
\begin{eqnarray}
\Gamma \lineup= \lambda \frac{1}{1+K}e^{\beta X^+}\frac{1}{1+K}\ -\ \lambda^2\frac{1}{1+K}e^{\beta X^+}\frac{1}{1+K}e^{\beta X^+}\frac{1}{1+K}\nonumber\\
\lineup\ \ \ \ \ \ \ \ \ \ \ \ \ \ \ \ \ \ \ \ \ \ \ \ \ \ \ \  \ +\ \lambda^3 \frac{1}{1+K}e^{\beta X^+}\frac{1}{1+K}e^{\beta X^+}\frac{1}{1+K}e^{\beta X^+}\frac{1}{1+K}\ -\ ... \ \ \ \ \ \  (\mathrm{early\ time}),\nonumber\\
\label{eq:gamma_initial}\\
\Gamma\lineup = \frac{1}{1+K} \ -\  \lambda^{-1} e^{-\beta X^+}\  +\  \lambda^{-2} e^{-\beta X^+}(1+K)e^{-\beta X^+}\nonumber\\
\lineup\ \ \ \ \ \ \ \ \ \ \ \ \ \ \ \ \ \ \ \ \ \ \ \ \ \ \ -\ \lambda^{-3} e^{-\beta X^+}(1+K)e^{-\beta X^+}(1+K)e^{-\beta X^+}\ + \ ... \ \ \ \ \ \ \ \ \ \ \ \ \ \ \ \ \  (\mathrm{late\ time}).\nonumber\\
\label{eq:gamma_final}
\end{eqnarray}
At first glance, the late time expansion appears to be somewhat singular. Each correction is given by a finite number of worldsheet operators acting on the identity string field. Usually such states would not be normalizable, but in the current situation there is no problem. Since $e^{-\beta X^+}$ is a matter operator of conformal dimension $-1$, observables are unambiguously defined in the presence of these corrections~\cite{IdSing}. 

\begin{figure}
\begin{center}
\resizebox{2.4in}{2.5in}{\includegraphics{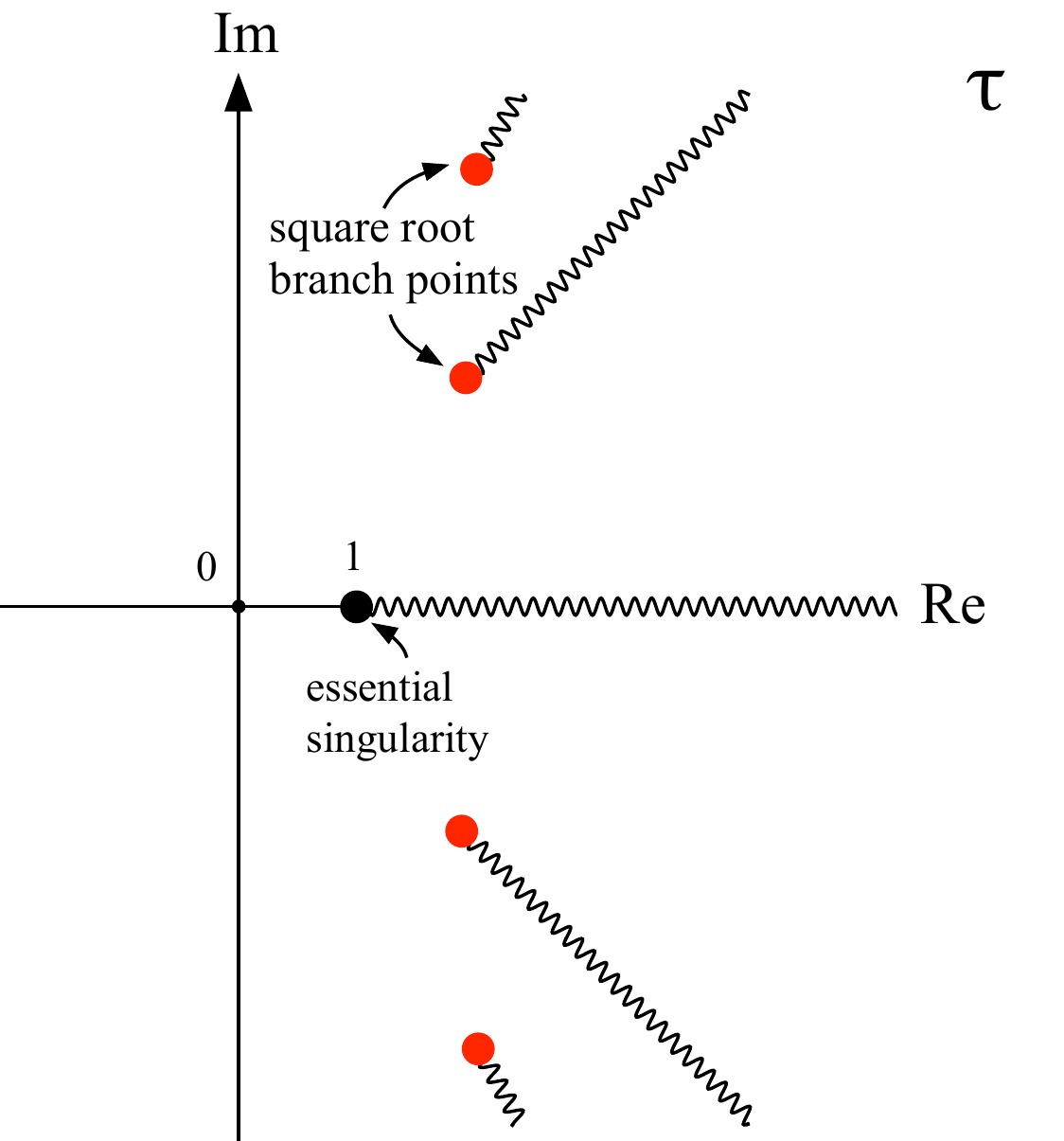}}
\end{center}
\caption{\label{fig:Borel} Singularities and branch cuts in the Borel plane for the ghost number zero tachyon. Since we will be interested in Borel summation, it is convenient to orient the branch cuts so that they extend directly away from the origin.}
\end{figure}

The central question we want to address is the range of validity of the late time expansion around the tachyon vacuum. We propose to understand this in terms of convergence of the coefficients in the basis of $L_0$ eigenstates for a given momentum. We discuss specifically the ghost number $0$ tachyon $\gamma(x)$. Let us define 
\begin{equation}\alpha \equiv \frac{8\lambda}{\pi}e^{\beta x^+},\end{equation}
which is a conveniently normalized expansion parameter. At early times $\gamma(x)$ can be written
\begin{equation}\gamma(x) = -\sum_{n=1}^\infty \frac{1}{n!}(-\alpha)^n\int_0^\infty dt\, e^{-t}\tau(t)^n.\label{eq:gammaintinitial}
\end{equation}
This series has infinite radius of convergence in $e^{\beta x^+}$. To see this, note that $|\tau(t)|\leq 1$ for all positive $t$, which implies
\begin{equation}\left|\int_0^\infty dt\, e^{-t}\tau(t)^n\right|\leq \int_0^\infty dt\, e^{-t} = 1,\end{equation}
and the series converges by comparison to the Taylor series for $e^{-\alpha}$. Since the expansion around $\alpha=0$ and $\alpha=\infty$ cannot both converge for the same $\alpha$, this implies that the late time expansion must have vanishing radius of convergence.  However, it is useful to understand this more directly. Suppose we change the integration variable from the Schwinger parameter $t$ to the Borel plane $\tau$:
\begin{equation}\gamma(x)=\alpha \int_0^1 d\tau\, e^{-t(\tau)}e^{-\alpha\tau},\label{eq:Borelgammaint}\end{equation}
where $t(\tau)$ is the inverse function of $\tau(t)$. The integration in the Borel plane ranges from $0$ to $1$. The late time expansion of this integral follows from Watson's lemma \cite{Miller}, which amounts to expanding $e^{-t(\tau)}$ in powers of $\tau$, integrating each term in the power series, and ignoring exponentially small contributions from the upper limit of integration. This gives:
\begin{equation}\gamma(x) \sim \sum_{n=0}^\infty \left.\frac{d^n}{d\tau^n}e^{-t(\tau)}\right|_{\tau=0} \left(\frac{1}{\alpha}\right)^n.\label{eq:gammaintfinal}\end{equation}
The growth of the coefficients is determined by the analytic structure of the integrand in the Borel plane. Since we do not have a closed form expression for $t(\tau)$, we have to figure this out indirectly. First, there will be a singularity for any finite $\tau$ which is the image of $|t|=\infty$. It is easy to check that $t=\infty$ maps to $\tau=1$, and in this vicinity we have an essential singularity with branch point:
\begin{equation}
e^{-t(\tau)}\approx \exp\left[-\frac{\pi}{2\sqrt{3}}\frac{1}{\sqrt{1-\tau}}\right],\ \ \  \tau\to 1.
\end{equation}
Second, there will be singularities for any $\tau$ which is the image of a saddle point $t$ of the function~$\tau(t)$. All of the saddle points are simple, so they produce square root branch points in the Borel plane. The singularities are located at 
\begin{equation}\tau = \frac{x_n}{2}\cot\frac{x_n}{2}\ \ \ \ n=1,2,3...,\label{eq:sqsingularity} \end{equation}
as well as at complex conjugate positions, where $x_n$ are roots of the equation
\begin{equation}
x_n=\sin x_n
\end{equation}
with positive real and imaginary parts increasing sequentially with $n$. A decent approximation of these roots, which deviates at most by a few percent for small $n$, is given by
\begin{equation}
x_n \approx \frac{\pi}{2}(4n+1)- \frac{2}{\pi}\frac{\ln(\pi(4n+1))}{4n+1}+i\ln(\pi(4n+1)).
\end{equation}
Plugging in the first few values of $x_n$, we find branch points at 
\begin{eqnarray}
\tau = 1.895 + 3.719 i, \ \ \ \  2.180 + 6.933 i, \ \ \ \  2.361 + 10.107 i,
\end{eqnarray}
as well as at the complex conjugate positions. These results are summarized in figure \ref{fig:Borel}. The growth of the coefficients in \eq{gammaintfinal} is determined by the singularity with closest proximity to the origin. Apparently, this is the essential singularity at $\tau=1$. Therefore the coefficients take the form
\begin{equation} \left.\frac{d^n}{d\tau^n}e^{-t(\tau)}\right|_{\tau=0}= n! f_n,\label{eq:growth}\end{equation}
where the numbers $f_n$ have slower than exponential decay for large $n$.\footnote{We have found that the $f_n$s are quite accurately approximated by the Taylor coefficients $\widetilde{f}_n$ of the function $\exp[-\frac{\pi}{2\sqrt{3}}(\frac{1}{\sqrt{1-\tau}}-1)]$ expanded around $\tau=0$. The asymptotics of $\widetilde{f}_n$ are fairly straightforward to derive by saddle point analysis of the Cauchy residue integral. We find 
\begin{equation} f_n\approx\widetilde{f}_n\sim -2\sqrt{2\pi}\left(\frac{\pi^2}{162 n^5}\right)^{1/6}e^{-\frac{1}{2}\left(\frac{3\pi}{4}\right)^{2/3}n^{1/3}-\frac{\pi}{2\sqrt{3}}}\sin\left(\frac{\sqrt{3}}{2}\left(\frac{3\pi}{4}\right)^{2/3}n^{1/3}-\frac{\pi}{6}\right)\label{eq:ass_guess}\end{equation} The coefficients decay exponentially with the cube root of $n$ and periodically change sign.} This establishes that the late time expansion has vanishing radius of convergence. We give a closed form expression for the $f_n$s in appendix \ref{app:f_n}.

Let us make a few comments:
\begin{itemize}
\item It is fortunate that the late time expansion has zero radius of convergence, since otherwise the solution would be physically equivalent to the tachyon vacuum at finite time---which, as explained in the introduction, would be problematic. However, vanishing radius of convergence is not quite sufficient to avoid this difficulty. It is possible that the gauge transformation to the tachyon vacuum, defined by an asymptotic expansion at late times, can be resummed into a gauge transformation which is also valid at finite time. Fortunately this does not occur. We give a fuller explanation in section \ref{sec:gauge}.
\item Due to the essential singularity in the Borel plane at $\tau=1$, the late time expansion is not Borel resummable. Therefore the evolution of the system  towards the tachyon vacuum must receive contribution from nonperturbative effects. The growth of the coefficients implies that nonperturbative corrections can be expected in the form
\begin{equation}e^{-\alpha}=e^{-\frac{8\lambda}{\pi}e^{\beta x^+}}.\end{equation}
One might guess that nonperturbative corrections represent a physical fluctuation of the tachyon vacuum at late times. However, this is not obvious. We return to this question later.
\item We have shown that the late time behavior of $\gamma(x)$ can be represented by a power series in $e^{-\beta x^+}$ with vanishing radius of convergence. However, we have not shown that this power series is actually computed from the proposed late time expansion of the string field \eq{gamma_final}. This question is significant since the relation between \eq{gamma_final} and the true late time behavior of the string field was only really justified at a symbolic level. In fact, \eq{gamma_final} produces a rather different formula for the coefficients of the power series at late times. To agree with \eq{gammaintfinal}, we must have the relation
\begin{eqnarray}\lineup\!\!\!\!\!\!\!\!\!\! \left.\frac{d^n}{d\tau^n}e^{-t(\tau)}\right|_{\tau=0}=\nonumber\\
\lineup\!\!\!\!\!\!\!\!\!\!(-1)^n \!\left.\!\left(\!1\!-\!\frac{\d}{\d \eps_1}\!\right)\!...\!\left(\!1\!-\!\frac{\d}{\d \eps_{n-1}}\!\right)\!\prod_{i=1}^n\!\left(\!\frac{2(1+\eps_1+...+\eps_{n-1})}{\pi}\!\sin\frac{\pi(\frac{1}{2}+\eps_i+...+\eps_{n-1})}{1+\eps_1+...+\eps_{n-1}}\!\right)^{2}\right|_{\eps_1,...,\eps_{n-1}=0}.\nonumber\\
\end{eqnarray}
We have tested this equality out to $n=12$, at which point symbolic computation of the right hand side becomes slow. By explicit evaluation of the derivatives one should be able to recast the right hand side in the form \eq{B8}, though the combinatorics seems prohibitive.
\end{itemize}

\section{Rolling to the Wrong Side of the Potential}
\label{sec:wrong}

In this section we would like to understand what happens if the tachyon falls towards negative values, where the tachyon effective potential is unbounded from below. At first this seems like a detour, but it turns out to be a significant part of the story. 

\begin{figure}
\begin{center}
\resizebox{3in}{2in}{\includegraphics{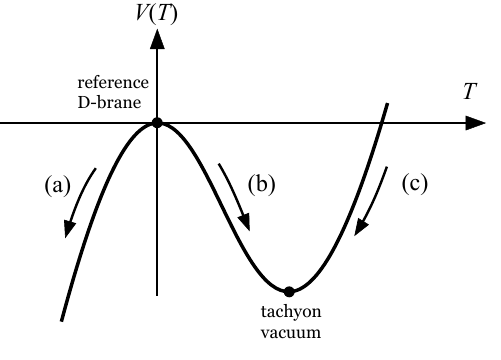}}
\end{center}
\caption{\label{fig:Potential} The light-like rolling tachyon solution can potentially describe three distinct time dependent backgrounds. Starting from the original D-brane configuration, case (a) represents the tachyon rolling towards negative values, where the effective potential is unbounded from below, and case (b) represents falling towards the tachyon vacuum. Case (c) represents rolling towards the tachyon vacuum from the ``other side'' on the tachyon effective potential.}
\end{figure}

Again we consider the ghost number $0$ factor in the solution
\begin{equation}
\Gamma = \frac{1}{1+K}-\frac{1}{1+K-|\lambda| e^{\beta X^+}},
\end{equation}
and assume $\lambda=-|\lambda|$ is negative, so the tachyon initially rolls towards negative values. Based on appearances, one might guess that the late time behavior of this state should be described by the asymptotic expansion \eq{gamma_final} with $\lambda<0$. This, however, would imply that the solution rolls to the tachyon vacuum even if the tachyon initially falls in the opposite direction. While this might be possible, it would be quite surprising. The situation is illustrated in figure \ref{fig:Potential}. The early time expansion for $\lambda<0$ has the tachyon falling to negative values, as in case (a); the late time expansion for $\lambda<0$, on the other hand, has the tachyon approaching the tachyon vacuum from larger expectation values, as in case (c). 

Since physics in the linear dilaton background is somewhat exotic, it is helpful to give some context by considering what happens in an ordinary scalar field theory model
\begin{equation}
S = -\int d^2 x\,e^{-V\cdot x} \left(\frac{1}{2}\partial^\mu\phi \partial_\mu\phi -\frac{1}{2}\phi^2 + \frac{1}{3}\phi^3\right),\label{eq:scalar}
\end{equation}
with conventional dilaton coupling and cubic potential. Assuming that $\phi$ is only a function of $x^+$, we can easily solve the equations of motion to find: 
\begin{equation}
\phi(x) = 1-\frac{1}{1+\lambda e^{\beta x^+}}.\label{eq:scalarsol}
\end{equation}
\noindent Surprisingly, this looks quite similar to the string field theory solution; formally, it is given by \eq{Gammaexp} after setting $K=0$. If $\lambda$ is negative, the scalar field initially falls to negative values as in case~(a), but begins to fall so rapidly that it reaches $-\infty$ in finite time. Thereafter we can continue to another solution which rolls from $+\infty$ at finite time down to the local minimum at $\phi=1$ in the infinite future, as in case (c). One might guess that something similar happens for the tachyon in string field theory. But this cannot be correct. The singularity in the evolution of $\phi(x)$ for $\lambda<0$ is possible since the early time expansion in powers of $e^{\beta x^+}$ has finite radius of convergence. But the early time expansion in string field theory has infinite radius of convergence. So a finite time singularity in the string field theory solution is not possible. 

However, the resemblance to the scalar field theory model can be enhanced with a different choice of gauge. Consider a solution of the form
\begin{equation}\Psi = \left[1-\frac{1}{1+\lambda e^{\beta X^+}}\right]c(1-K).\label{eq:idsol}\end{equation} 
The tachyon coefficient is given by
\begin{equation}T(x)= \frac{1}{2\pi}\gamma(x)(1+2\gamma(x)),\end{equation}
where $\gamma(x)$ is the ghost number zero tachyon coefficient of the factor in square brackets:
\begin{equation}\gamma(x)=1-\frac{1}{1+2\pi\lambda e^{\beta x^+}}.\end{equation}
For this solution, the ghost number zero tachyon evolves in exactly the same way as $\phi(x)$ in the scalar field theory model. For $\lambda<0$ the string field hits a singularity, and for $\lambda>0$ the late time expansion around the tachyon vacuum has finite radius of convergence, appearing to imply that the decay process ends at the tachyon vacuum in finite time. This unphysical behavior is present because the solution is singular from the perspective of the identity string field \cite{IdSing}---essentially, the solution does not contain enough ``worldsheet" to correctly capture the stringy physics of the decay process. We believe that regular string field theory solutions will not exhibit this unphysical behavior.\footnote{In \cite{Hellerman} it was shown that the $\mathcal{B}_0$ gauge solution approaches the tachyon vacuum when expanded in a basis of eigenstates of the operator
\begin{equation}\mathcal{L}_0-iV\cdot p
\end{equation} where $p$ is the momentum. The coefficient fields in this basis behave similarly to the solution of the scalar field theory model. However, the eigenstates of the basis are singular from the perspective of the identity string field. Effectively, the basis represents the $\mathcal{B}_0$ gauge solution as a perturbation of the singular solution \eq{idsol}, and the unphysical features of the singular solution persist to any finite order in the expansion.}

Returning to the original solution, for $\lambda<0$, the trajectory of the ghost number zero tachyon is given by the integral
\begin{equation}
\gamma(x) = 1-\int_0^\infty dt\, e^{-t+|\alpha|\tau(t)},\ \ \ \ (\alpha=-|\alpha|<0) .\label{eq:neglint}
\end{equation}
For late times, an upper bound and reasonable estimate of this integral can be found by replacing $\tau(t)$ by $1$ in the integrand, which leads to
\begin{equation} -e^{\frac{8|\lambda|}{\pi}e^{\beta x^+}}\lesssim\gamma(x).\end{equation}
This indicates that the tachyon keeps rolling to arbitrarily negative values for all time, and does not reach the tachyon vacuum. 

Let us improve this estimate. For large negative $\alpha$, the integral is dominated by a critical point (or saddle point from the point of view of the complex plane) where the growth of $e^{|\alpha| \tau(t)}$ is just being compensated by the decay of~$e^{-t}$. The location of the critical point can be determined by solving
\begin{equation}\frac{d}{dt}(t+\alpha\tau(t))=0.\end{equation}
If we define
\begin{equation}
x\equiv\frac{\pi}{t+1},
\end{equation}
this leads to a transcendental equation
\begin{equation}
\sin x -x= \frac{2\pi}{\alpha}\frac{1-\cos x}{x^2}.\label{eq:saddles}
\end{equation}
When applying the saddle point method, usually we would be justified in ignoring the right hand side of this equation, which is vanishingly small for large $\alpha$. Then the saddle points of the integrand are determined by the roots of $x=\sin x$:
\begin{equation}0,\ x_n,\ x_n^*,\ -x_n,\ -x_n^*,\label{eq:gammasaddle}\end{equation}
where the $x_n$s were introduced in the last section. All of these roots except $x=0$ represent saddle points clustered around an infinite collection of essential singularities at the positions
\begin{equation}t_\mathrm{singularity} = -1+\frac{1}{2n},\ \ \ \ n=\pm 1, \ \pm 2,\ \pm 3,...\ . \label{eq:tsingularity}\end{equation}

\begin{figure}
\begin{center}
\resizebox{3in}{2.3in}{\includegraphics{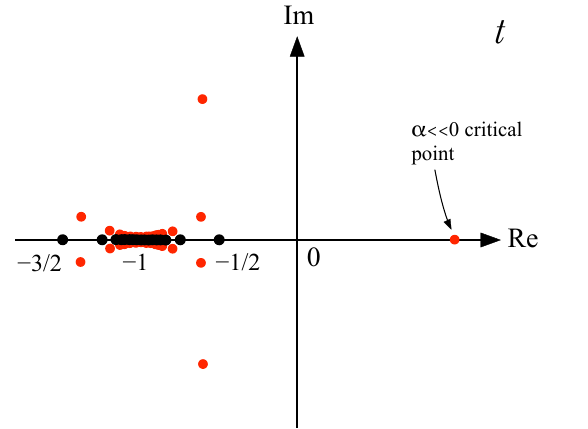}}
\end{center}
\caption{\label{fig:Schwinger} Location of saddle points and singularities in the integrand of the ghost number zero tachyon in the Schwinger plane. The black dots represent essential singularities related to poles of the cotangent. There are an infinite number of them, and they accumulate towards $t=-1$. The singularity with closest proximity to the origin is at $t=-1/2$. The red dots represent saddle points. The saddle points which are clustered around the essential singularities correspond to the roots $x_n,n\geq 1$ of the equation $x=\sin x$. In addition there are three saddle points, shown here for $\alpha< 0$, whose location approaches infinity as $|\alpha|$ becomes large. We show these saddle points for finite $\alpha$ so that they can be seen on the same scale as the singularities on the negative real axis. The saddle point on the positive  real axis dominates the behavior of the solution for $\alpha\ll 0$.}
\end{figure}

\noindent This is shown in figure~\ref{fig:Schwinger}.  These saddle points, however, are not directly relevant for the late time behavior. The critical point which interests us corresponds (roughly) to $x=0$.  We say ``roughly" because, in this case, we cannot ignore the right hand side of \eq{saddles} in determining the position of the corresponding saddle point. Solving \eq{saddles} for small $x$ perturbatively when $|\alpha|$ is large gives
\begin{equation}
t_\mathrm{saddle}=\left(\frac{\pi^2}{6}\right)^{1/3}(-\alpha)^{1/3}-1+\frac{1}{15}\left(\frac{\pi^2}{6}\right)^{2/3}(-\alpha)^{-1/3}+O(1/\alpha).\label{eq:tsaddle}
\end{equation}
This formula actually determines three saddle points, corresponding to three cube roots of unity. For large negative $\alpha$, the saddle point which defines the maximum of the integrand \eq{neglint} is real and positive, and increases with $x^+$ as the cube root of $-\alpha$. Approximating $\gamma(x)$ by a Gaussian integral centered on the critical point then gives the late time behavior 
\begin{equation}
\gamma(x) \sim -\left(\frac{4\pi^5}{81}\right)^{1/6}(-\alpha)^{1/6}\exp\left[-\alpha -\left(\frac{3\pi}{4}\right)^{2/3}(-\alpha)^{1/3}+1\right]\Big(1+O(\alpha^{-1/3})\Big).
,\ \ \ (\alpha\ll 0)\label{eq:saddle}\end{equation}
The right hand side is multiplied by an asymptotic series in inverse powers of $\alpha$. Ignoring subleading multiplicative factors, $\gamma(x)$ grows as $-e^{\frac{8|\lambda|}{\pi}e^{\beta x^+}}$ at late times, agreeing with our initial estimate. From this analysis, we learn two interesting facts. First, the solution does not reach the tachyon vacuum when the tachyon rolls in the wrong direction. In particular, the asymptotic expansion \eq{gamma_final} does not correctly describe the late time behavior when $\lambda<0$. Second, we note the following. When the tachyon rolls in the positive direction, towards the tachyon vacuum, the late time behavior is dominated by the region $t=0$ of the integrand, which represents the identity string field. The identity-like nature of the late time expansion is apparent in \eq{gamma_final}. By contrast, when the tachyon rolls in the negative direction, away from the tachyon vacuum, the late time behavior is dominated by large positive $t$. This represents the sliver state. 

\begin{figure}
\begin{center}
\resizebox{3.5in}{2.9in}{\includegraphics{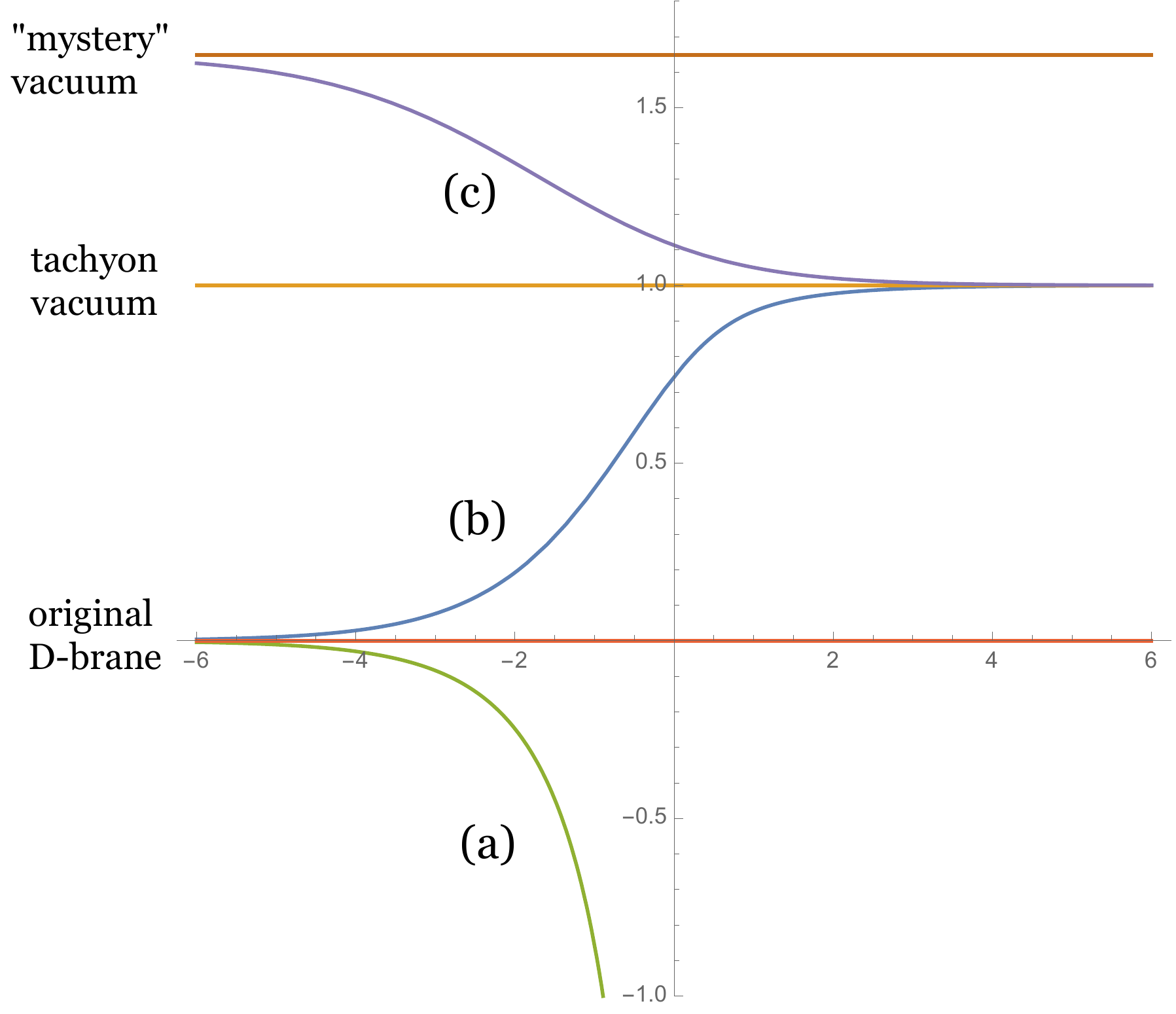}}
\end{center}
\caption{\label{fig:mystery} Profiles of the ghost number zero tachyon as a function of $x^+\in[-6,6]$ for three solutions: Case (a) is when the tachyon rolls in the negative direction,  case (b) is when the tachyon rolls towards the tachyon vacuum (already shown in figure \ref{fig:gamma}), and case (c) is a solution which rolls to the tachyon vacuum from the ``other side." The initial configuration for case (c) appears to represent a new vacuum state.}
\end{figure}

This leaves an open question: What is the meaning of the asymptotic expansion \eq{gamma_final} when $\lambda<0$? Apparently it represents a solution where the tachyon rolls ``backwards" to the tachyon vacuum from larger expectation values. Of course, \eq{gamma_final} only gives an asymptotic expansion of the solution, but we are lucky since for $\lambda<0$ the expansion is  Borel resummable. Using the coordinate $\tau$ on the Borel plane introduced before, resummation gives a formula for the ghost number $0$ tachyon: 
\begin{equation}
\gamma(x) = |\alpha|\int_{-\infty}^0 d\tau \, e^{-t(\tau)+|\alpha|\tau}.
\end{equation}
The integration is performed on the negative real axis since $\alpha$ is negative. Transforming to the Schwinger parameterization, we can equivalently write
\begin{equation}
\gamma(x) = 1+\int_{-1/2}^0 dt\, e^{-t+|\alpha|\tau(t)}. \label{eq:gammac}
\end{equation}
Now comes a surprise. The resummed solution appears to represent decay from a new vacuum state which sits on the ``other side" of the local minimum of the tachyon effective potential. At the new vacuum the ghost number zero tachyon takes the value
\begin{equation}\lim_{x^+\to-\infty}\gamma(x)=\sqrt{e}>1.\end{equation}
We plot the corresponding profile of $\gamma(x)$ in figure \ref{fig:mystery}.

What is the meaning of this solution? From the range of integration over the Schwinger parameter in \eq{gammac}, it is clear that Borel summation gives an unusual definition to the second term in $\Gamma$:
\begin{equation}
\Gamma= \frac{1}{1+K}-\sigma\frac{1}{1+K}\overline{\sigma}.
\end{equation}
Apparently, the second term should be defined by
\begin{equation}
\frac{1}{1+K} = -\int_{-1/2}^0 dt\, e^{-t}\Omega^t.\label{eq:weirdinv}
\end{equation}
Since $t$ is negative, the solution involves wedge states with negative width. Such states are highly non-normalizable; component field expectation values grow without bound as we ascend the infinite tower of states towards increasing $L_0$ eigenvalue. Therefore it seems that the asymptotic series for $\lambda<0$ does not resum into a well-behaved state. Still, we do not wish to disregard the solution entirely; it may represent something important that---for whatever reason---is difficult for the string field algebra to capture. However, the expansion into a basis of $L_0$ eigenstates does not give a satisfactory definition of the solution; star products computed component-wise in the $L_0$ basis fail to converge. We circumvent this difficulty with the following prescription. Suppose we want to compute a number-valued functional of the solution $F[\Psi]$. The asymptotic expansion for $\lambda<0$ allows us to compute this functional at least as a formal power series in $1/\lambda$:
\begin{equation}F[\Psi]\sim\sum_{n=0}^\infty \frac{1}{\lambda^n}F_n.\label{eq:resumll0}\end{equation}
We then propose to define $F[\Psi]$ through Borel summation of this power series. This is in fact how we derived the formula for the ghost number $0$ tachyon from the beginning.

Let us address one further question, which helps to understand the significance of the complicated singularity/saddle point structure in the Schwinger plane for $\mathrm{Re}(t)<0$. The question is why the Schwinger integration in the solution \eq{weirdinv} ends at $t=-1/2$. As suggested by notation, $\frac{1}{1+K}$ should define a star algebra inverse for $1+K$,
\begin{equation}(1+K)\frac{1}{1+K}\equiv 1,\end{equation}
but when the Schwinger integration ends at $t=-1/2$, instead we have 
\begin{equation}(1+K)\frac{1}{1+K} = 1-\sqrt{e}\,\Omega^{-1/2}.\end{equation}
The boundary term at $t=-1/2$ does not vanish. However, it {\it does} vanish---in a limited sense we now describe---when appearing between $\sigma$ and $\overline{\sigma}$: 
\begin{equation}\sigma \Omega^{-1/2}\overline{\sigma}=0.\end{equation}
To see why, consider contracting $\sigma \Omega^t \overline{\sigma}$ with an $L_0$ eigenstate in the form of a plane wave vertex operator inserted on the $SL(2,\mathbb{R})$ vacuum. This can be computed as a correlation function on the cylinder:
\begin{equation}
\Tr\Big[\sigma\Omega^t\overline{\sigma}\sqrt{\Omega}e^{ik\cdot X}\sqrt{\Omega}\Big]^{X^0,X^1}=\left\langle\exp\left[|\lambda|\int_{1/2}^{1/2+t}dz\, e^{\beta X^+}(z)\right]e^{ik\cdot X}(0)\right\rangle_{C_{t+1}}^{X^0,X^1}.
\end{equation} 
The sign in front of the integral in the exponential is positive, since we are now concerned with $\lambda<0$. When the wedge width $t$ is negative, the upper limit of integration is smaller than the lower limit, and we can interchange the limits to produce a minus sign:
\begin{equation}
\Tr\Big[\sigma\Omega^t\overline{\sigma}\sqrt{\Omega}e^{ik\cdot X}\sqrt{\Omega}\Big]^{X^0,X^1}=\left\langle\exp\left[-|\lambda|\int_{1/2-|t|}^{1/2}dz\, e^{\beta X^+}(z)\right]e^{ik\cdot X}(0)\right\rangle_{C_{1-|t|}}^{X^0,X^1}, \ \ \ \ \ \ (-1/2<t<0).
\end{equation}
As we approach $t=-1/2$ the upper and lower limits of the integration collide with the plane wave vertex operator (recall that on a cylinder $C_L$, coordinates are identified $z\sim z+L$). This produces a divergence from the OPE between $e^{\beta X^+}$ and $e^{ik\cdot X}$, and the sign in front of the integral in the exponential is such that this divergence makes the correlation function vanish. In fact, this is the reason why the integrand of $\gamma(x)$ has an essential singularity at $t=-1/2$. Continuing further to $t<-1/2$, the boundary deformation will repeatedly wind around the cylinder as the circumference shrinks to zero, colliding with the probe vertex operator when
\begin{equation}\frac{1}{2}=0\ \mathrm{mod}\ 1-|t|.\end{equation}
This precisely corresponds to the location of the essential singularities \eq{tsingularity} in the Schwinger plane for $\mathrm{Re}(t)<0$. Therefore we find that \eq{weirdinv} does in a sense define an inverse for $1+K$, at least when sandwiched between $\sigma$ and $\overline{\sigma}$. However, it is clear that the mechanism behind this depends heavily on the assumption that we expand in a basis of $L_0$ eigenstates---the singularity $t=-1/2$ is directly related to the $\sqrt{\Omega}$ which appears when $L_0$ eigenstates are expressed in the sliver frame. This indicates that the complicated singularity/saddle point structure of the integrand for $\mathrm{Re}(t)<0$ probably does not have physical significance, at least in details.

\section{Nonperturbative Effects}
\label{sec:Stokes}

We are now ready to describe the nonperturbative contributions to the late time decay process. We focus on the ghost number $0$ tachyon
\begin{equation}\gamma(x) = 1-\int_0^\infty dt \,e^{-t-\alpha \tau(t)}.\end{equation}
Nonperturbative effects are related to nontrivial saddle points of the ``action functional"
\begin{equation}t+\alpha\tau(t).\end{equation}
When the tachyon falls to negative values ($\lambda<0$), we have seen that the late time behavior is dominated by a sliver-like saddle point \eq{tsaddle}. It is natural to guess that this saddle point is also responsible for nonperturbative corrections when the solution rolls to the tachyon vacuum. To understand this, we consider complex~$\lambda$ ,
\begin{equation}\lambda = -|\lambda|e^{i\theta},\label{eq:complexa}\end{equation}
and track the contribution of the saddle point to the late time behavior as $\theta$ ranges from $0$ to $\pi$ (the story for $\theta\in[-\pi,0]$ is simply obtained by complex conjugation). Because of the sign in \eq{complexa}, $\theta=0$ represents rolling in the negative direction, and $\theta=\pi$ represents rolling towards the tachyon vacuum. The saddle point contribution can be found using the method of steepest descent. We deform the integration contour in the complex Schwinger plane,
\begin{equation}t\in[0,\infty],\end{equation}
into a homotopically equivalent curve consisting of segments with the property that
\begin{equation}\mathrm{Im}(t+\alpha\tau(t))\end{equation}
is constant. In the limit of large $|\alpha|$, integration along each segment will produce its own characteristic asymptotic behavior, and the asymptotic behavior of the entire integral is determined by the segment which produces the dominant contribution. 

\begin{figure}
\begin{center}
\resizebox{3.2in}{2.4in}{\includegraphics{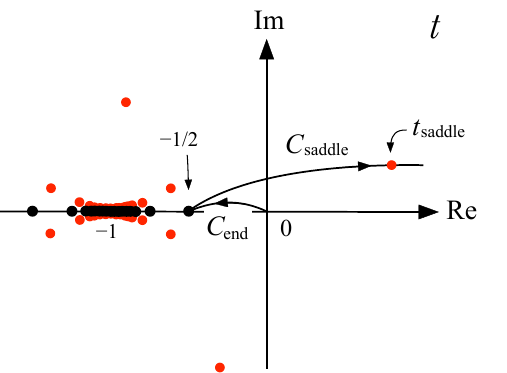}}
\end{center}
\caption{\label{fig:steepest1} Components of the steepest descent contour of the ghost number zero tachyon with $\lambda =-|\lambda| e^{i \theta}$ and $0\leq\theta<\pi/2$. In this range of angles, $C_\mathrm{saddle}$ makes the dominant contribution to the asymptotic behavior. We plot the contour for finite $\alpha$ so that the saddle point $t_\mathrm{saddle}$ can be seen on the same scale as the singularities on the negative real axis.}
\end{figure}

Let us first assume that the angle $\theta$ is in the range $0\leq\theta<\pi/2$. In this case, the contour of steepest descent is composed of two segments. The first segment is a contour $C_\mathrm{end}$ emanating from the origin on the path of steepest descent characterized by
\begin{equation}\mathrm{Im}(t+\alpha\tau(t))=0,\ \ \ t\in C_\mathrm{end}.\end{equation}
This connects $t=0$ with the essential singularity at $t=-1/2$. The second segment is a contour $C_\mathrm{saddle}$ which passes through the saddle point \eq{tsaddle}. If we take $t_\mathrm{saddle}$ to be the root with the smallest angle relative to the positive real axis, we have
\begin{equation}\mathrm{Im}(t+\alpha\tau(t))=\mathrm{Im}(t_\mathrm{saddle}+\alpha\tau(t_\mathrm{saddle})),\ \ \ t\in C_\mathrm{saddle},\end{equation}

\begin{figure}
\begin{center}
\resizebox{3.2in}{2.4in}{\includegraphics{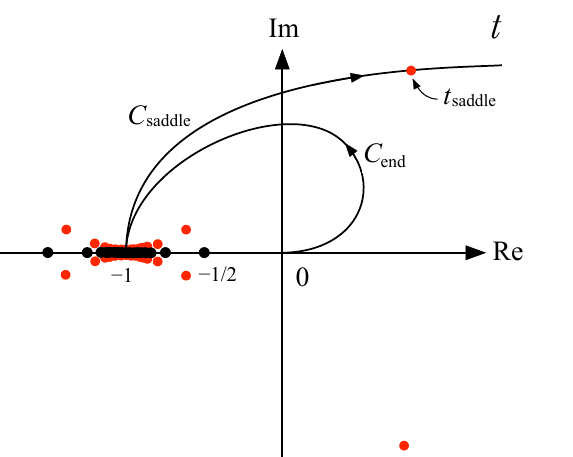}}
\end{center}
\caption{\label{fig:steepest2} Steepest descent contour for the ghost number zero tachyon with $\lambda =-|\lambda| e^{i \theta}$ and $(.650...)\pi<\theta<\pi$. We plot the contour for finite $\alpha$ so that the saddle point $t_\mathrm{saddle}$ can be seen on the same scale as the singularities on the negative real axis.}
\end{figure}

\noindent which connects the essential singularity at $t=-1/2$ to $t=\infty$. The two components of the total contour are shown in figure \ref{fig:steepest1}. Deforming the standard  path $t\in[0,\infty]$ we can write
\begin{equation}\gamma(x) = 1-\int_{C_\mathrm{end}}dt\,e^{-t-\alpha \tau(t)}-\int_{C_\mathrm{saddle}}dt\,e^{-t-\alpha \tau(t)}.\end{equation}
For $0\leq\theta< \pi/2$, the integration over $C_\mathrm{saddle}$ produces the dominant asymptotic behavior \eq{saddle} (now with complex $\alpha$). The integration over $C_\mathrm{end}$ is insignificant by comparison. Interestingly, however, $C_\mathrm{end}$ gives precisely the asymptotic expansion around the tachyon vacuum for $\mathrm{Re}(\lambda)< 0$. Therefore, the mysterious solution which rolls to the tachyon vacuum from the ``other side" is responsible for a negligible, but distinguished contribution to the late time behavior when the tachyon rolls towards negative values.

As we reach $\theta=\pi/2$ the contribution from $C_\mathrm{saddle}$ turns from exponentially dominant to exponentially suppressed. The $C_\mathrm{end}$ contour then becomes more important. In the standard terminology, $\theta=\pi/2$ represents an anti-Stokes line. Continuing to larger angles, after a brief transition,\footnote{For angles $\pi/2<\theta<(.650...)\pi$ the contours $C_\mathrm{end}$ and $C_\mathrm{saddle}$ end on different essential singularities, and it is necessary to include contours passing through other saddle points to get something homotopically equivalent to the $t\in[0,\infty]$. This is a manifestation of the Stokes phenomenon arising from the square root branch point singularities \eq{sqsingularity} in the Borel plane. The contribution from the other contours in this range of angles is subdominant to $C_\mathrm{end}$. Since this phenomenon is far from the physical angles $\theta=0,\pi$, we will not analyze it further.} the steepest descent contour again has two segments $C_\mathrm{end}$ and $C_\mathrm{saddle}$ as shown in figure \ref{fig:steepest2}. The intersection point of $C_\mathrm{end}$ and $C_\mathrm{saddle}$ now sits on the essential singularity at $t=-1$. As we approach $\theta=\pi$, the contour $C_\mathrm{end}$ unravels to cover the entire positive real axis and the vertical line $\mathrm{Re}(t)=-1,\mathrm{Im}(t)>0$. The significance of this is more easily understood in the Borel plane, where $\gamma(x)$ can be written as
\begin{equation}\gamma(x) = \alpha\int_{\tau\circ C_\mathrm{end}}d\tau \,e^{-t(\tau)-\alpha\tau}+\alpha\int_{\tau\circ C_\mathrm{saddle}}d\tau\, e^{-t(\tau)-\alpha\tau}.\end{equation}
As $\theta$ approaches $\pi$, the contour $\tau\circ C_\mathrm{end}$ approaches the positive real axis from above. This represents an (upper) lateral Borel sum of the late time asymptotic series for $\lambda>0$. Meanwhile, the contour $\tau\circ C_\mathrm{saddle}$ connects $\tau=+\infty$ to the essential singularity at $\tau=1$. This partially cancels the integration of the lateral Borel transform, so that at $\theta=\pi$ we obtain
\begin{equation}\gamma(x) = \alpha\int_0^1 d\tau \,e^{-t(\tau)-\alpha\tau},\end{equation}
in accordance with \eq{Borelgammaint}. In the context of resurgence theory, nonperturbative corrections can be identified with whatever contributions should be added to a (lateral) Borel transform of an asymptotic series in order to obtain the correct nonperturbative result. Therefore, we have shown that the nonperturbative corrections are given precisely by integration over the steepest descent contour $C_\mathrm{saddle}$ passing through the saddle point $t_\mathrm{saddle}$.

\begin{figure}
\begin{center}
\resizebox{2.5in}{2.4in}{\includegraphics{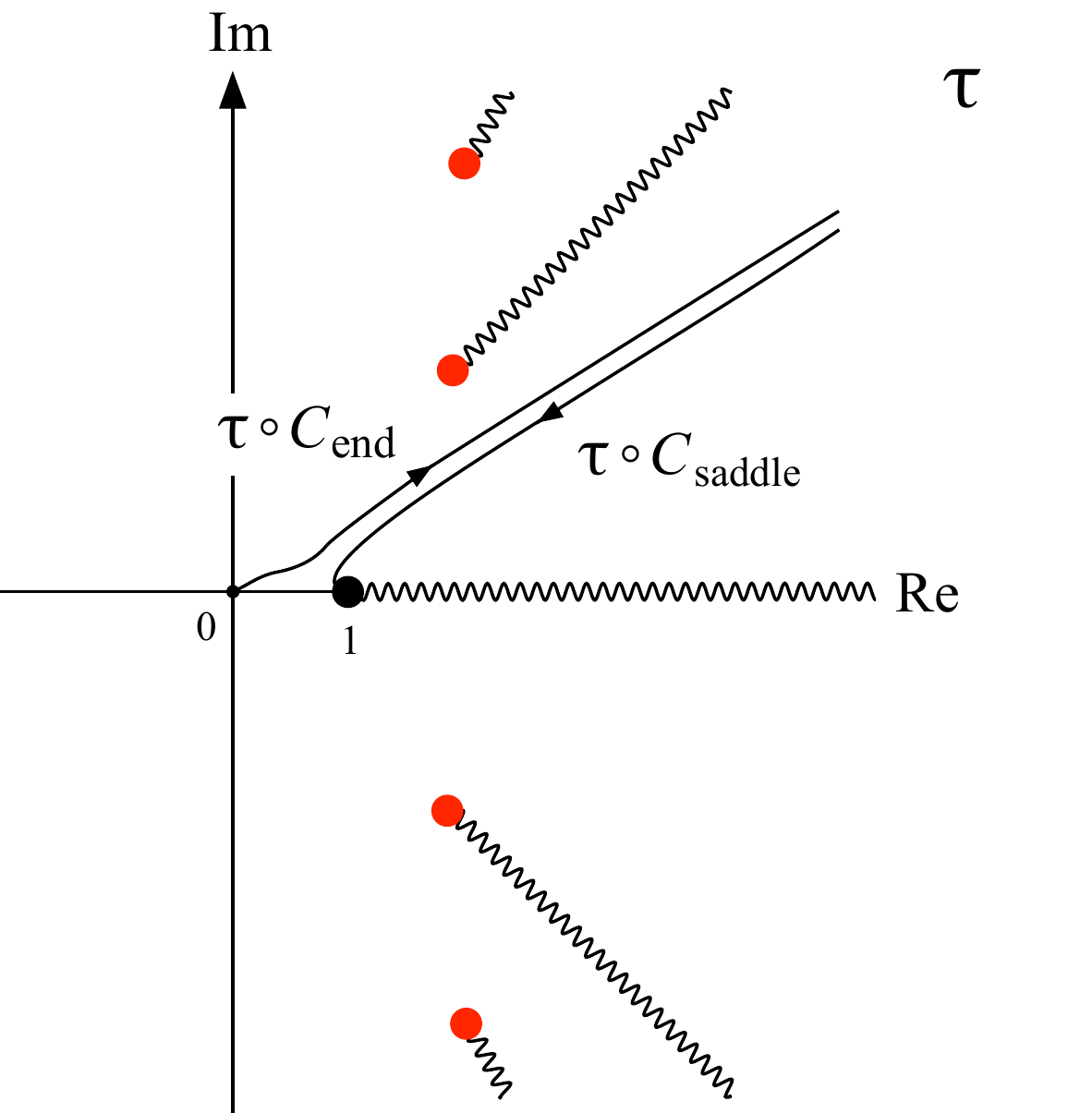}}
\end{center}
\caption{\label{fig:steepest3} Image of the steepest descent contours $C_\mathrm{end}$ and $C_\mathrm{saddle}$ in the Borel plane as $\theta$~approaches $\pi$ from below, for large (but not infinite) $\alpha$. In the limit $\alpha\to\infty$, the image of $C_\mathrm{end}$ becomes a straight line connecting the origin to infinity at an angle of $\pi-\theta$.}
\end{figure}

As we cross $\theta=\pi$ the steepest descent contours suddenly shift: the relevant root $t_\mathrm{saddle}$ and the contours $C_\mathrm{end}$ and $C_\mathrm{saddle}$ are reflected about the real axis relative to $0<\theta<\pi$. This means that $\theta=\pi$ represents a Stokes line. As $\theta$ approaches $\pi$ from larger angles, the requisite nonperturbative corrections are complex conjugate to those when $\theta$ approaches $\pi$ from smaller angles. This is expected, since the difference in the appropriate saddle point contributions is required to cancel the difference between upper and lower lateral Borel transforms of the asymptotic series, giving the same nonperturbative result. This is explained more fully in section \ref{sec:resurgence}.

\section{Physical Interpretation of Nonperturbative Effects}
\label{sec:physical}

Since the perturbative fluctuations of the tachyon vacuum at late time are pure gauge, it is natural to infer that any physical fluctuation of the tachyon vacuum is related to nonperturbative effects. However, it seems difficult to make this intuition precise, since any fluctuation of the tachyon vacuum presumably disappears in the infinite future, and in a theory of quantum gravity we do not expect the finite time configuration to be observable. However, the situation is not as bad as one might think; we are dealing with purely classical open string theory, and gravity is not present. In fact, we will argue that the finite time configuration can at least formally be probed with a purely on-shell amplitude.

We consider the amplitude for emission of a single closed string from the decaying D-brane:
\begin{equation}
F(k)=\left\langle \mathcal{V}(0,0) c(1) \exp\left[-\lambda\int_{0}^{2\pi} d\theta\, e^{\beta X^+}(e^{i\theta})\right]\right\rangle_\mathrm{disk},
\label{eq:amp}\end{equation}
where $\mathcal{V}(z,\overline{z})$ is an on-shell closed string vertex operator, and the nonlocal exponential transforms the initially Neumann boundary condition into the time dependent boundary condition characterizing the light-like rolling tachyon background. We want to choose a closed string vertex operator which allows us to say something about the state of the system at finite time. This can be accomplished by the following trick \cite{Kawano,boundarystate}. We assume that, in addition to $X^0,X^1$, there is another noncompact, space-like free boson $Y(z,\overline{z})$ orthogonal to the D-brane subject to Dirichlet boundary conditions. We can choose a closed string vertex operator which carries arbitrary momentum $k$ along the ``interesting" directions $X^0,X^1$, while along the $Y$ direction it carries a momentum $p$ which is tuned so as to stay on the mass shell. We consider specifically the vertex operator for the closed string tachyon
\begin{equation}
\mathcal{V}(z,\overline{z}) = -c\overline{c}\,e^{ipY} e^{-i(k+2i V)\cdot X}(z,\overline{z}).\label{eq:V}
\end{equation}
The bulk plane wave vertex operator $e^{ik\cdot X}(z,\overline{z})$ is a spinless primary of dimension $\frac{1}{2}k^2+iV\cdot k$. The mass shell condition amounts to the requirement that $\mathcal{V}(z,\overline{z})$ is a spinless primary operator of dimension $0$, and this fixes the momentum along the $Y$ direction to be given by 
\begin{equation}p^2 = 4 -k^2-2iV\cdot k.\end{equation}
The main observation is that the $Y$ momentum of the vertex operator does not contribute to the amplitude; the 1-point function of $e^{ip Y}$ on the disk  is equal to unity,
\begin{equation}\langle e^{ipY}(0,0)\rangle_\mathrm{disk}^Y = 1,\end{equation}
thanks to the Dirichlet boundary condition. This means that when computing the tadpole amplitude we can drop the $e^{ip Y}$ factor, and we are effectively probing the decay process with an off-shell vertex operator. This allows us to extract gauge invariant information about the state of the system at finite time.  
In fact, we have chosen $\mathcal{V}(z,\overline{z})$ so that the amplitude $F(k)$ represents the leading coefficient of the boundary state in an expansion into a basis of $L_0+\overline{L}_0$ eigenstates for fixed momentum:
\begin{equation}
|B\rangle = \int\frac{d^2 k dp}{(2\pi)^3}F(k)(c_0+\overline{c}_0)c_1\overline{c}_1e^{i k\cdot X}e^{i p Y}(0,0)|0\rangle+...,
\end{equation}
where $|0\rangle $ is the $SL(2,\mathbb{C})$ vacuum of the bulk CFT and
\begin{equation}
F(k)= \langle \mathcal{V},c_0^-B\rangle,
\end{equation}
where the closed string inner product is defined with $c_0^-=\frac{1}{2}(c_0-\overline{c}_0)$. It is worth noting that the amplitude has delta function support on a discrete set of momenta satisfying
\begin{equation} k = -iV +in\beta \delta^+,\end{equation}
where $n$ represents the order $\lambda^n$ contribution to the exponential operator in \eq{amp}. The required momentum $p$ along the $Y$ direction is given by
\begin{equation}p=2i\sqrt{n-1}.\end{equation}
Since the momenta are imaginary, the contributing vertex operators do not correspond to normalizable closed string asymptotic states. We will not speculate as to what this implies about the practical possibility of measurement; we only regard the amplitude as observable in a formal sense.

\begin{figure}
\begin{center}
\resizebox{3in}{1.8in}{\includegraphics{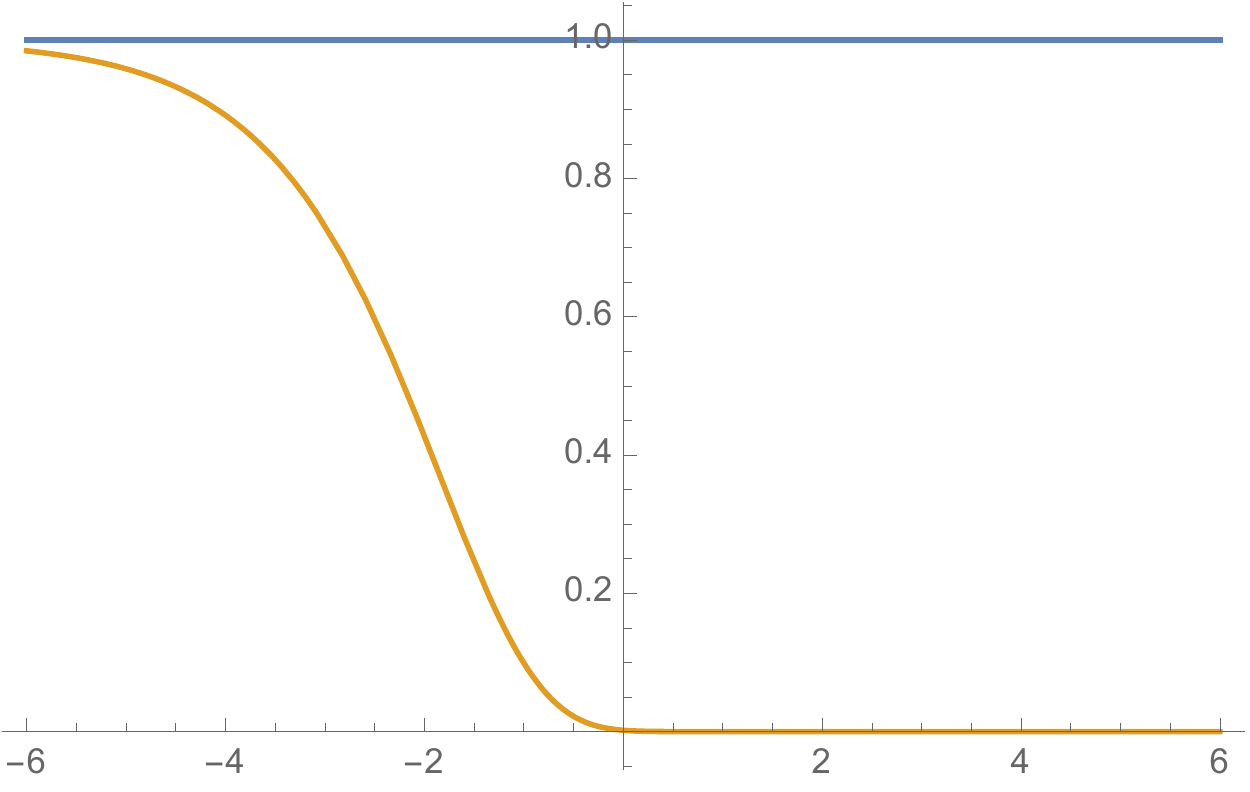}}
\end{center}
\caption{\label{fig:tadpole} Tadpole amplitude $F(x)$ as a function of $x^+\in[-6,6]$ with $x^-=0$. }
\end{figure}

We are interested in the position space profile of the amplitude, obtained through the Fourier transform
\begin{equation}F(x) = \int\frac{d^2 k}{(2\pi)^2}e^{ik\cdot x}F(k).\end{equation}
Following appendix \ref{app:bcc} one can show that
\begin{equation}F(x) = e^{-\frac{1}{\beta}x^-}\exp(-2\pi \lambda e^{\beta x^+}).\label{eq:phys_amp}\end{equation}
The $x^-$ dependent factor is present for all coefficients of the boundary state and reflects the fact that the strength of the D-brane backreaction increases towards the strong coupling region. The interesting part is the $x^+$ dependent factor, which says that initially there is a steady source for closed strings which rapidly disappears near $x^+=0$, as shown in figure \ref{fig:tadpole}. Afterwards the amplitude is super-exponentially small but not quite zero; there is a tiny but persistent residual source for closed strings. This means that the background does not reach the tachyon vacuum in finite time. 

The tadpole amplitude can be directly related to the string field theory solution $\Psi$ using the Ellwood invariant \cite{tadpole}:
\begin{equation}
F(x) = -2\pi i\int\frac{d^2k}{(2\pi)^2} e^{ik\cdot x}\Tr_\mathcal{V}[\Psi-\Psi_\mathrm{tv}],
\end{equation}
where $\Psi_\mathrm{tv}$ is a solution for the tachyon vacuum, which we can conveniently choose to be the first term in \eq{KOS}. We can ask how the late time expansion of the solution contributes to the amplitude. Using standard manipulations we arrive at
\begin{eqnarray}
F(x) \lineup = 2\pi i\int\frac{d^2k}{(2\pi)^2}e^{ik\cdot x}\Tr_\mathcal{V}\left[\frac{1}{1+K+\lambda e^{\beta X^+}}Bc\d c\right]\nonumber\\
\lineup \sim 2\pi i\int\frac{d^2k}{(2\pi)^2}e^{ik\cdot x}\left(\frac{1}{\lambda}\Tr_\mathcal{V}[e^{-\beta X^+}Bc\d c]-\frac{1}{\lambda^2}\Tr_\mathcal{V}[e^{-\beta X^+}(1+K)e^{-\beta X^+}Bc\d c]+\ ...\ \right).\nonumber\\
\end{eqnarray}
The ``traces" which appear at each order  in $1/\lambda$ represent correlation functions on a degenerate cylinder with vanishing circumference, but each cylinder contains operator insertions with net negative scaling dimension. Such correlation functions vanish identically \cite{IdSing}. Therefore the late time asymptotic expansion contributes nothing to the amplitude:
\begin{equation}
F(x) \sim 2\pi i\left(\frac{1}{\lambda}0-\frac{1}{\lambda^2}0+\frac{1}{\lambda^3}0-....\right).
\end{equation}
This fits with the expectation that the perturbative fluctuations around the tachyon vacuum are pure gauge. But note that Borel summation of a vanishing power series will give a vanishing result. This means that the entire amplitude must be generated by nonperturbative effects.

However, the late time behavior of the tadpole amplitude is too degenerate to make the meaning of this statement clear; the asymptotic expansion vanishes identically, there is no singularity in the Borel plane, and no obvious sense in talking about contributions from nontrivial saddle points. To clarify this, we consider a deformation of the amplitude given by replacing the implicit overlap with the identity string field in the Ellwood invariant with an overlap with a wedge state of positive width $\Delta$:
\begin{equation}
\widetilde{F}(x,\Delta) = -2\pi i\int\frac{d^2k}{(2\pi)^2} e^{ik\cdot x}\Tr_\mathcal{V}[\Omega^\Delta(\Psi-\Psi_\mathrm{tv})].
\end{equation}
This ``amplitude" is not gauge invariant. This is by design, since otherwise the late time asymptotic expansion of the amplitude would vanish identically. However, the amplitude becomes physical in the limit $\Delta\to0$. Plugging in the solution we find two contributions
\begin{equation}
\widetilde{F}(x,\Delta) = -2\pi i \int\frac{d^2 k}{(2\pi)^2}e^{ik\cdot x}\bigg(\Tr_\mathcal{V}\left[\sigma\frac{1}{1+K}\overline{\sigma} \Omega^\Delta Bc\d c\right]-\underbrace{\Tr_\mathcal{V}\left[\sigma\frac{1}{1+K}\overline{\sigma}(1+K)Bc\Omega^\Delta  c\right]}_{\mathrm{ignore}}\bigg).
\end{equation}
We will henceforth ignore the second term since it vanishes in the limit $\Delta\to 0$ (due to $c^2=0$), and cannot make a physical contribution to nonperturbative effects. We denote the first term as $F(x,\Delta)$. We can compute using standard manipulations to find
\begin{equation}F(x,\Delta) = e^{-\frac{1}{\beta}x^-}\int_0^\infty dt\, e^{-t}\exp\left[-2\pi\lambda e^{\beta x^+}\frac{t}{t+\Delta}\right].\label{eq:FDelta}\end{equation}
The integration variable $t$ is the Schwinger parameter for the expansion of $1/(1+K)$ into a superposition of wedge states. When $\Delta=0$ the integration over $t$ is trivial and we recover the physical amplitude \eq{phys_amp}.

The asymptotic analysis of $F(x,\Delta)$ is  analogous to that of the ghost number zero tachyon, but simpler. The naturally normalized expansion parameter  is
\begin{equation}\alpha = 2\pi \lambda e^{\beta x^+}.\end{equation}
Note that $\alpha$ is similar to the expansion parameter of the ghost number zero tachyon (also called~$\alpha$), but with different normalization. The integrand in the Schwinger plane has one essential singularity~at
\begin{equation}t_\mathrm{singularity} =-\Delta,\end{equation}
and a pair of distant saddle points at positions
\begin{equation}t_\mathrm{saddle}=-\Delta\pm\sqrt{-\alpha\Delta}.\label{eq:Ftsingularity}\end{equation}
There is a clear connection between the essential singularity at $-\Delta$ and the essential singularity at $-1$ in the integrand of the ghost number zero tachyon. Both result from a negative Schwinger parameter implying a correlation function on a cylinder of vanishing circumference. 
The pair of saddle points \eq{Ftsingularity} should also be analogous to the triplet of saddle points \eq{tsaddle} in the integrand of the ghost number zero tachyon. However, the integrand of $F(x,\Delta)$ does not have an analogue of the infinite cluster of essential singularities \eq{tsingularity} and associated saddle points \eq{gammasaddle}. This is because the boundary condition changing operators do not collide with the probe vertex operator for negative $t$, since the probe vertex operator is in the bulk. We can transform the integrand from the Schwinger plane~$t$ to the Borel plane $\tau$:
\begin{equation}\tau = \frac{t}{t+\Delta},\ \ \ \ t=\frac{\Delta \tau}{1-\tau}.\end{equation}
Conveniently, the transformation has an inverse in closed form. The amplitude expressed in the Borel plane is 
\begin{equation}F(x,\Delta) = e^{-\frac{1}{\beta}x^-}\int_0^1d\tau \frac{\Delta}{(1-\tau)^2}\exp\left[-\frac{\Delta\tau}{1-\tau}\right]e^{-\alpha\tau}.\label{eq:FBorel}\end{equation}
The integrand has an essential singularity (without branch point) at $\tau=1$. This implies that the late time asymptotic expansion of $F(x,\Delta)$ is not Borel resummable, and receives corrections from nonperturbative effects. In fact, we can express the late time expansion in closed form; in the Borel integrand we recognize the generating function for associated Laguerre polynomials. This leads to
\begin{equation}
F(x,\Delta)\sim \Delta e^{-\frac{1}{\beta}x^-}\sum_{n=0}^\infty n! L_n^{(1)}(\Delta)\frac{1}{\alpha^{n+1}}.\label{eq:Fintfinal}
\end{equation}
The entire asymptotic series is multiplied by $\Delta$. In the physical limit $\Delta\to 0$ the asymptotic expansion vanishes, as expected. 

\begin{figure}
\begin{center}
\resizebox{2.8in}{2.5in}{\includegraphics{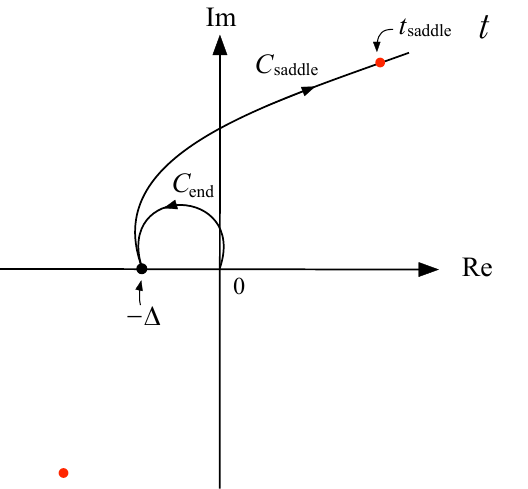}}\ \ \ \ \ 
\resizebox{2.4in}{2.5in}{\includegraphics{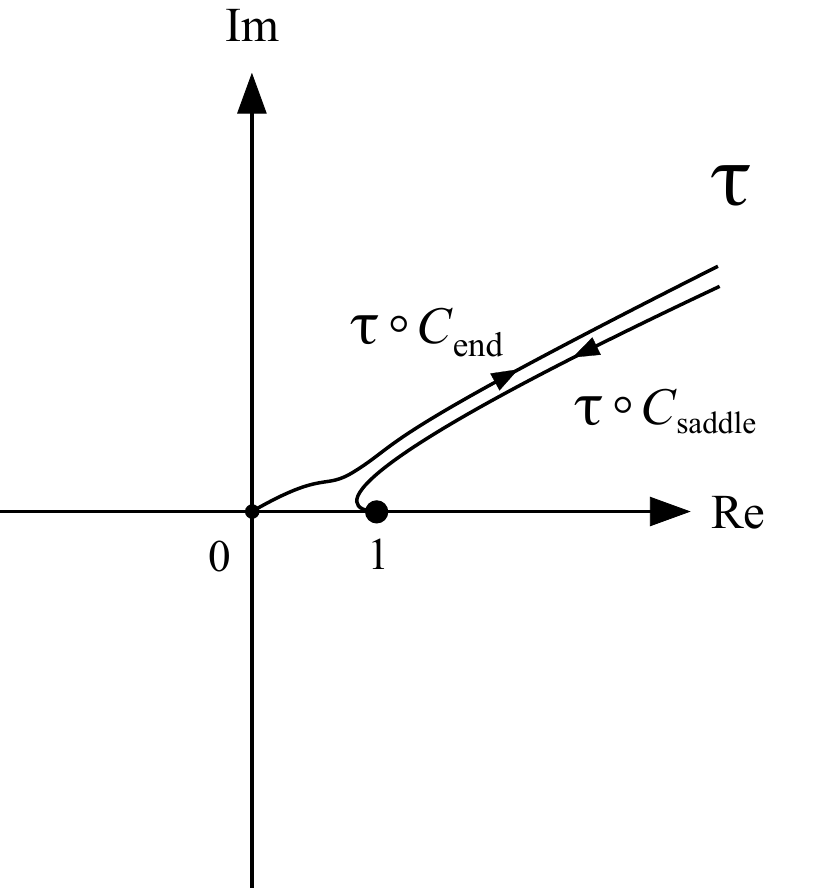}}
\end{center}
\caption{\label{fig:Fsteepest} Steepest descent contours $C_\mathrm{end}$ and $C_\mathrm{saddle}$ of the integrand of the amplitude $F(x,\Delta)$, shown left in the Schwinger plane $t$, and right in the Borel plane $\tau$.}
\end{figure}

To understand the contribution of nonperturbative effects, we revisit the steepest descent analysis. We consider complex $\lambda$,
\begin{equation}\lambda = -|\lambda|e^{i\theta},\end{equation}
and track the saddle point contribution to the amplitude as $\theta$ varies from $0$ to $\pi$. As in section~\ref{sec:Stokes}, $\theta=0$ corresponds to rolling towards the unbounded side of the tachyon effective potential and $\theta=\pi$ corresponds to rolling towards the tachyon vacuum. Similarly to the ghost number zero tachyon, the amplitude can be written as an integral over an ``endpoint" steepest descent contour $C_\mathrm{end}$ and a ``saddle point" steepest descent contour $C_\mathrm{saddle}$:
\begin{equation}F(x,\Delta) =  e^{-\frac{1}{\beta}x^-}\left(\int_{C_\mathrm{end}} dt\, e^{-t}\exp\left[-\alpha\frac{t}{t+\Delta}\right]+\int_{C_\mathrm{saddle}}dt\, e^{-t}\exp\left[-\alpha\frac{t}{t+\Delta}\right]\right),\end{equation}
where
\begin{eqnarray}
\mathrm{Im}\left(t+\alpha\frac{t}{t+\Delta}\right)\lineup =0,\ \ \ t\in C_\mathrm{end},\nonumber\\
\mathrm{Im}\left(t+\alpha\frac{t}{t+\Delta}\right)\lineup =\mathrm{Im}\left(t_\mathrm{saddle}+\alpha\frac{t_\mathrm{saddle}}{t_\mathrm{saddle}+\Delta}\right),\ \ \ t\in C_\mathrm{saddle},
\end{eqnarray}
where $t_\mathrm{saddle}$ is the saddle point \eq{Ftsingularity} with the smallest angle relative to the positive real axis. The endpoint contour $C_\mathrm{end}$ connects the origin $t=0$ to the essential singularity at $t=-\Delta$, and the saddle point contour $C_\mathrm{saddle}$ passes from $t=-\Delta$, through the saddle point $t=t_\mathrm{saddle}$, and on to $t=\infty$. This is shown in figure \ref{fig:Fsteepest}. For $0\leq\theta<\pi/2$, the saddle point contour makes the dominant contribution, representing super-exponential growth as the tachyon rolls towards negative values. At $\theta=\pi/2$ we encounter an anti-Stokes line, and the saddle point contribution turns from exponentially dominant to exponentially suppressed. For larger angles $\pi/2<\theta<\pi$ the contour $C_\mathrm{end}$ and the associated asymptotic expansion \eq{Fintfinal} gives the dominant contribution to the late time behavior. As we approach $\theta=\pi$, it is more convenient to visualize the steepest descent contours in the Borel plane. The image of the endpoint contour $\tau\circ C_\mathrm{end}$ extends from the origin at an angle $\pi-\theta$ and passes to $\tau=\infty$, and the image of the saddle point contour $\tau\circ C_\mathrm{saddle}$ passes from $\tau=\infty$, swings around and approaches the essential singularity at $\tau=1$ from below. This is shown in figure \ref{fig:Fsteepest}. Near $\theta=\pi$, the contour $\tau\circ C_\mathrm{end}$ represents the (upper) lateral Borel transform of the asymptotic series \eq{Fintfinal}. The integration over $\tau\circ C_\mathrm{saddle}$ partially cancels that of the lateral Borel transform, producing the expected integration of $\tau$ from $0$ to $1$. This shows that the nonperturbative corrections to the late time expansion of $F(x,\Delta)$ are given precisely by integration over the steepest descent contour $C_\mathrm{saddle}$. The analysis follows quite closely what we have already seen for the ghost number zero tachyon.

We can now meaningfully talk about the contribution of nonperturbative effects to the physical amplitude in the limit $\Delta\to 0$. In this limit, the saddle point contour $C_\mathrm{saddle}$ approaches the positive real axis in the limit $\Delta\to 0$ for fixed $x^+$. Meanwhile, the nontrivial dependence on $t$ in the integrand disappears:
\begin{equation}\lim_{\Delta\to 0}\exp\left[-\alpha\frac{t}{t+\Delta}\right]=e^{-\alpha}.\end{equation}
The integration over $C_\mathrm{saddle}$ then becomes trivial, and we recover precisely the physical amplitude~\eq{phys_amp}. In summary, we have shown that it is possible to meaningfully talk about a physical fluctuation of the tachyon vacuum at late times, and this fluctuation is described in open string field theory by nonperturbative corrections to the late time asymptotic expansion of the solution.

Let us revisit the puzzling solution for $\lambda<0$ which appears to roll ``backwards" to the tachyon vacuum from larger tachyon expectation values. Following the prescription discussed at the end of section \ref{sec:wrong}, we can compute the amplitude $F(x,\Delta)$ of this solution by Borel summation of the asymptotic series \eq{Fintfinal} for $\lambda<0$. This gives
\begin{equation}
F(x,\Delta) = \Delta e^{-\frac{1}{\beta}x^-}\int_{-\infty}^0 d\tau \frac{1}{1-\tau}\exp\left[-\frac{\Delta \tau}{1-\tau}\right]e^{|\alpha|\tau}.
\end{equation}
Note that $\Delta$ multiplies an integral which is finite as $\Delta$ approaches zero. Therefore the amplitude vanishes in the physical limit. This is directly related to the fact that the solution---as it is defined by Borel summation---does not receive correction from nonperturbative effects. Therefore the solution must be pure gauge. This is quite interesting, since it lends support to the idea that there is no physics on the ``other side" of the local minumum of the tachyon effective potential. For example, in the ``exact" tachyon effective potential of boundary string field theory \cite{Shatashvilli,Moore},
\begin{equation}V(T)\propto (1+T)e^{-T},\end{equation}
the tachyon vacuum sits at $T=\infty$, and it is not meaningful to ask what lies beyond. Our analysis can be viewed as evidence that this is a genuine physical statement, and not an artifact of some singular field redefinition.

\section{Resurgence}
\label{sec:resurgence}

In this section we describe our results from the point of view of the theory of resurgence \cite{Ecalle} (see \cite{Unsal,Marino,Schiappa} for reviews). This formalism is perhaps more than is needed for present purposes; a major motivation of resurgence is to reconstruct the nonperturbative answer from perturbation theory around saddle points. But we already have the nonperturbative answer: it is given by the exact solution \eq{KOS}. However, resurgence gives some insight into why the nonperturbative solution takes the form that it does; in a more complicated setup, it is possible that a string field theory solution will only be known through perturbative expansions, and the connection to resurgence theory could be useful. 

We consider the regularized tadpole amplitude \eq{FDelta} instead of the ghost number zero tachyon, since the function theory here is more tractable. The question is how much we can learn about the exact amplitude from the perturbative expansions around the tachyon vacuum and the saddle points \eq{Ftsingularity}. The idea is to represent the amplitude $F(x,\Delta)\equiv F(\alpha)$ through a trans-series, which we denote as $f(\alpha)$,
\begin{equation}
f(\alpha) = \sigma_0(\theta)f_\mathrm{end}(\alpha)+\sigma_+(\theta)f_\mathrm{saddle}^+(\alpha)+\sigma_-(\theta)f^-_\mathrm{saddle}(\alpha). \label{eq:trans}\end{equation}
where $f_\mathrm{end}(\alpha)$ is the perturbative expansion around the tachyon vacuum \eq{Fintfinal} and $f_\mathrm{saddle}^\pm(\alpha)$ are the expansions around the saddle points \eq{Ftsingularity} with respectively positive or negative real part.  The coefficients $\sigma_0(\theta),\sigma_\pm(\theta)$ are called trans-series parameters, and are piecewise constant functions of the argument of $-\alpha = |\alpha| e^{i\theta}$. It is assumed that the trans-series parameters can only be discontinuous for angles $\theta$ where the trans-series is not Borel resummable. The perturbative expansions are known or will soon be derived, so the primary objective is to determine the trans-series parameters. This is done by requiring that Borel summation of the trans-series defines a nonperturbative expression for $F(\alpha)$ which is unambiguously defined both when rolling to the tachyon vacuum ($\alpha>0$ and $\theta=\pi$) and when rolling away from the tachyon vacuum, where the tachyon effective potential is unbounded from below ($\alpha<0$ and $\theta=0$).

The first step is to derive the perturbative expansions around the saddle points. Together with an additive shift by $-\Delta$, the saddle points are defined by 
\begin{eqnarray}
t_+\lineup \equiv e^{i\theta/2}\sqrt{|\alpha|\Delta},\\
t_- \lineup \equiv -e^{i\theta/2}\sqrt{|\alpha|\Delta},
\end{eqnarray}
where $-\pi<\theta<\pi$. Note that $t_+$ represents the saddle point with positive real part, while $t_-$ represents the saddle point with negative real part. The saddle point $-\Delta+t_-$ did not play a role in previous analysis; we will see why in a moment. But without prior knowledge, we must account for the possibility that $t_-$ will also give some contribution to nonperturbative effects. As we approach $\theta = \pm \pi$ the two saddle points approach the vertical line $\mathrm{Re}(t)=-\Delta$,  and $t_+$ crosses over into $t_-$, and vice-versa: 
\begin{equation}t_+|_{\theta=\pi}=t_-|_{\theta=-\pi}\ \ \ \ t_+|_{\theta=-\pi}=t_-|_{\theta=\pi}.\end{equation}
The perturbative expansions $f_\mathrm{saddle}^\pm(\alpha)$ are defined by the asymptotic behavior of the integral 
\begin{equation}F^\pm_\mathrm{saddle}(\alpha) = \int_{C^\pm_\mathrm{saddle}} dt \exp\left[-t-\alpha\frac{t}{t+\Delta}\right]e^{-x^-/\beta},\label{eq:Fsaddleint}\end{equation}
where $C^\pm_\mathrm{saddle}$ is the steepest descent contour passing from the essential singularity at $-\Delta$, through the respective saddle point at $-\Delta + t_\pm$, and on to infinity. For large $|\alpha|$, the integral can be approximated by expanding around the saddle point
\begin{equation}t = -\Delta +t_{\pm}+\delta t,\end{equation}
and treating cubic and higher order couplings of $\delta t$ as small. This gives
\begin{eqnarray}
f_\mathrm{saddle}^\pm(\alpha) \lineup = \int_{\mathrm{through}\ \delta t=0} d(\delta t)\exp\left[-\alpha -2t_\pm +\Delta -\frac{1}{t_\pm}(\delta t^2)+\sum_{n=3}^\infty\frac{(\delta t)^n}{(-t_\pm)^{n-1}}\right]e^{-x^-/\beta},\label{eq:fsaddleint}
\end{eqnarray}
where the contour passes through the saddle point in the direction of increasing $|t|$, at an angle such that the quadratic term in $\delta t$ is real and negative; the precise form of the contour away from the saddle point is irrelevant for the large $|\alpha|$ expansion. The leading order contribution for large $|\alpha|$ is given by performing the Gaussian integral: 
\begin{equation}f_\mathrm{saddle}^\pm(\alpha) = \sqrt{\pi t_\pm}e^{-\alpha-2t_\pm+\Delta }\Big(1+\mathcal{O}(t_\pm^{-1})\Big)e^{-x^-/\beta},\end{equation}
where the square root is defined using
\begin{eqnarray}\mathrm{arg}(t_+)\lineup =\frac{\theta}{2},\\
\mathrm{arg}(t_-)\lineup = \frac{\theta}{2}-\pi\,\sgn(\theta).\end{eqnarray}
The discontinuity in $\mathrm{arg}(t_-)$ at $\theta=0$ is implied by the orientation of the steepest descent contour, and also ensures that the perturbative expansions $f_\mathrm{saddle}^+$ and $f_\mathrm{saddle}^-$ match as we approach $\theta=\pm\pi$. The factor
\begin{equation}e^{-\alpha-2t_\pm+\Delta }\end{equation}
is the exponential of the ``action" evaluated on the saddle point. Its presence reflects the fact that $f^\pm_\mathrm{saddle}$ is a nonperturbative effect. This is multiplied by an asymptotic series in powers of $\frac{1}{t_\pm}\sim\frac{1}{\sqrt{\alpha}}$. We can derive the asymptotic series by accounting for the contributions of the cubic and higher order terms in \eq{fsaddleint}. The easiest way to do this is simply to realize that the integral \eq{Fsaddleint} can be carried out exactly:
\begin{equation}F^\pm_\mathrm{saddle}(\alpha) = 2 t_\pm e^{-\alpha+\Delta}K_1(2 t_\pm)e^{-x^-/\beta}.\label{eq:Fpmexact}\end{equation}
Then we can use the known formula for the asymptotic expansion of the modified Bessel function. This gives the complete formula for the perturbative expansions around the saddle points:
\begin{equation}
f_\mathrm{saddle}^\pm (\alpha)=\frac{4}{\sqrt{\pi}} t_\pm^{3/2}e^{-\alpha-2t_\pm+\Delta}\left(\sum_{n=0}^\infty(-1)^{n+1}\frac{\Gamma(n-\frac{1}{2})\Gamma(n+\frac{3}{2})}{\Gamma(n+1)}\frac{1}{(4 t_\pm)^{n+1}}\right)e^{-x^-/\beta}.
\end{equation}
Since there are more factorial factors in the numerator than in the denominator, the expansion is asymptotic.

To recover the exact amplitude $F(\alpha)$, we need to apply Borel summation to the trans-series~\eq{trans}. Let us describe our conventions for Borel summation. Suppose we are given an expansion of the form
\begin{equation}a(z) = N(z)p(c z^\nu),\end{equation}
where $c,\nu$ are constants, $p(x)$ is a formal power series 
\begin{equation}p(x) = \sum_{n=0}^\infty \frac{p_n}{x^{n+1}},\end{equation}
and $N(z)$ is a function of $z$. Typically, $p(x)$ will be an asymptotic series, and $N(z)$ will not have an expansion in inverse powers of $z$. The Borel transform of $p(x)$ is defined by summing the power series
\begin{equation}\mathcal{B}[p](\xi) = \sum_{n=0}^\infty\frac{p_n}{n!}\xi^n,\end{equation}
followed by analytic continuation in the Borel plane $\xi$. Borel summation of $a(z)$ is then defined~by
\begin{equation}
\mathcal{S}[a](z) = N(z)\int_0^{\,+\infty\times e^{-i\,\mathrm{arg}(c z^\nu)}} d\xi\, e^{-(c z^\nu) \xi}\,\mathcal{B}[p](\xi).
\end{equation}
The integration in the Borel plane extends from the origin to infinity along a ray which ensures that $(c z^\nu) \xi$ is positive. If $\mathcal{B}[p](\xi)$ does not have singularities along the ray or blow up too fast at infinity, the integral will converge. With this definition, Borel summation of the expansion around the tachyon vacuum \eq{Fintfinal} gives
\begin{equation}
F_\mathrm{end}(\alpha) \equiv \mathcal{S}[f_\mathrm{end}](\alpha) = \Delta \int_0^{-\infty\times e^{-i\theta}} d\tau\, e^{-\alpha\tau}\frac{1}{(1-\tau)^2}\exp\left[-\frac{\Delta\tau}{1-\tau}\right] e^{-x^-/\beta}.
\end{equation}
The resummation is ambiguous at $\theta = \pm \pi$ due to the essential singularity at $\tau = 1$, as noted in the previous section. Meanwhile, Borel summation of the expansion around the saddle points gives
\begin{equation}
F_\mathrm{saddle}^\pm (\alpha)=\mathcal{S}[f_\mathrm{saddle}^\pm](\alpha)=4\sqrt{\pi} t_\pm^{3/2}\,e^{-\alpha-2t_\pm+\Delta}\int_0^{\pm\infty\times e^{i\theta/2}} d\xi \, e^{-4 t_\pm\xi} \!\phantom{a}_2 F_1\left(-\frac{1}{2},\frac{3}{2},1,-\xi\right)e^{-x^-/\beta}.
\end{equation}
Evaluating the integral reproduces the formula \eq{Fpmexact} in terms of the modified Bessel function. The resummation is ambiguous at $\theta=\pm\pi$, since the corresponding contours in the Borel plane do not match. Additionally, $F_\mathrm{saddle}^-$ is ambiguous at $\theta = 0$ due to a branch point from the hypergeometric function at $\xi=-1$. Therefore, the full amplitude after resummation,
\begin{equation}F(\alpha) = \sigma_0(\theta)F_\mathrm{end}(\alpha)+\sigma_+(\theta)F_\mathrm{saddle}^+(\alpha) + \sigma_-(\theta)F_\mathrm{saddle}^-(\alpha),\end{equation}
is generically ambiguous for real and physical $\alpha$. When the decay evolves towards the tachyon vacuum, an ambiguity is present in all three terms; when the decay evolves towards the side of the tachyon effective potential which is unbounded from below, an ambiguity appears from $F_\mathrm{saddle}^-$. 

The idea is to fix the trans-series parameters so that these ambiguities cancel. Consider first the ambiguity at $\theta=0$. We require that
\begin{equation}F(-|\alpha|e^{i0^+}) = F(-|\alpha|e^{i0^-}).\end{equation}
Since $F_\mathrm{end}$ and $F_\mathrm{saddle}^+$ are continuous through $\theta=0$, this condition implies
\begin{eqnarray}
0 \lineup = \Big(\sigma_0(0^+)-\sigma_0(0^-)\Big)F_\mathrm{end}(-|\alpha|)+\Big(\sigma_+(0^+)-\sigma_+(0^-)\Big)F_\mathrm{saddle}^+(-|\alpha|)\nonumber\\
\lineup\ \ \ +\sigma_-(0^+)F_\mathrm{saddle}^-(-|\alpha|e^{i0^+}) - \sigma_-(0^-)F_\mathrm{saddle}^-(-|\alpha|e^{i0^-}). 
\end{eqnarray}
The functions appearing on the right hand side are linearly independent, so equality will only hold if their coefficients are zero. This implies that $\sigma_0,\sigma_+,\sigma_-$ are continuous through $\theta=0$, and further that $\sigma_-$ vanishes at $\theta=0$. By assumption, the only other place where the trans-series parameters may be discontinuous is at $\theta=\pm\pi$, but since a piecewise constant function on a circle can only be discontinuous at an even number of points, this implies that the trans-series parameters must be independent of $\theta$:
\begin{equation}\sigma_0(\theta) = 1,\ \ \ \sigma_+(\theta) = \sigma_+,\ \ \ \sigma_-(\theta)=0.\end{equation}
Therefore, the $t^-$ saddle point cannot contribute to the amplitude. Next, we require that the amplitude is unambiguous at $\theta=\pm\pi$:
\begin{equation}F(-|\alpha|e^{i \pi^-})=F(-|\alpha| e^{-i\pi^-}).\end{equation}
This implies
\begin{equation}
0=F_\mathrm{end}(-|\alpha|e^{i \pi^-})-F_\mathrm{end}(-|\alpha|e^{-i\pi^-})+\sigma_+\Big(F_\mathrm{saddle}^+(-|\alpha|e^{i \pi^-})-F_\mathrm{saddle}^+(-|\alpha|e^{-i \pi^-})\Big).
\end{equation}
Now the ambiguity of the nonperturbative saddle point contribution must cancel the Borel ambiguity of the expansion around the tachyon vacuum. The former is given by the residue of the essential singularity at $\tau=1$:
\begin{eqnarray}
F_\mathrm{end}(-|\alpha|e^{i \pi^-})-F_\mathrm{end}(-|\alpha|e^{-i\pi^-})\lineup = -\Delta\oint_{\tau=1}d\tau e^{-|\alpha|\tau}\frac{1}{(1-\tau)^2}\exp\left[-\frac{\Delta\tau}{1-\tau}\right]e^{-x^-/\beta}\nonumber\\
\lineup = 2 \pi i\sqrt{|\alpha|\Delta}e^{-|\alpha|+\Delta} J_1(2\sqrt{|\alpha|\Delta})e^{-x^-/\beta}\phantom{\Bigg)}.
\end{eqnarray}
Using \eq{Fpmexact} and the identity
\begin{equation}K_1(ix) +K_1(-ix)=-\pi J_1(x),\end{equation}
the ambiguities cancel when 
\begin{equation}\sigma_+=1.\end{equation}
Therefore all trans-series parameters have been fixed, and the nonperturbative amplitude is given~by
\begin{eqnarray}F(\alpha)\lineup = F_\mathrm{end}(\alpha)+F_\mathrm{saddle}^+(\alpha)\nonumber\\
\lineup =  e^{-\frac{1}{\beta}x^-}\int_0^1d\tau \frac{\Delta}{(1-\tau)^2}\exp\left[-\frac{\Delta\tau}{1-\tau}\right]e^{-\alpha\tau},\end{eqnarray}
which is the correct result.

\section{Gauge Transformation to the Tachyon Vacuum}
\label{sec:gauge}

There is a general expectation that any solution which is parametrically close to the tachyon vacuum should be gauge equivalent to the tachyon vacuum. We would now like to explain why this expectation fails for the light-like rolling tachyon solution.

Consider a tachyon vacuum solution $\Psi_\mathrm{tv}$. A distinguishing property of $\Psi_\mathrm{tv}$ is the existence of a string field $A$ of ghost number $-1$, called the homotopy operator, satisfying
\begin{equation}Q_{\Psi_\mathrm{tv}}A = 1,\end{equation}
where $Q_{\Psi_\mathrm{tv}} = Q+[\Psi_{\mathrm{tv}},\cdot]$ is the shifted kinetic operator around the tachyon vacuum. The existence of the homotopy operator implies that $Q_{\Psi_\mathrm{tv}}$ has no cohomology; at the linearized level, all fluctuations around the tachyon vacuum are pure gauge. With some qualification, this statement also extends to the nonlinear equations of motion. Given a generic solution $\Psi$, consider the state
\begin{equation}
U = 1+(\Psi-\Psi_\mathrm{tv})A.\label{eq:U}
\end{equation}
This state satisfies
\begin{equation}
(Q+\Psi)U = U\Psi_\mathrm{tv}.\label{eq:almostgt}
\end{equation}
If $U$ has an inverse, this implies that $\Psi$ is gauge equivalent to $\Psi_\mathrm{tv}$: 
\begin{equation}
U^{-1}(Q+\Psi)U = \Psi_\mathrm{tv}.
\end{equation}
Generally $U$ will not have an inverse. However, if $\Psi$ is parametrically close to $\Psi_\mathrm{tv}$, it should be possible to construct an inverse perturbatively. This is why one doesn't expect to find nontrivial solutions close to the tachyon vacuum.

Let us see how this works for the light-like rolling tachyon solution. For the tachyon vacuum of \cite{simple} the homotopy operator takes the form
\begin{equation}A=\frac{B}{1+K},\end{equation}
and plugging into \eq{U} gives 
\begin{equation}
U = 1-cB(1+K)\sigma\frac{1}{1+K}\overline{\sigma}\frac{1}{1+K}.\label{eq:Unon}
\end{equation}
At late times this can be expanded perturbatively in powers of $e^{-\beta X^+}$:
\begin{eqnarray}
U \lineup = 1-cB(1+K)\left[\frac{1}{\lambda}e^{-\beta X^+} -\frac{1}{\lambda^2}e^{-\beta X^+}(1+K)e^{-\beta X^+}\right.\nonumber\\
\lineup\ \ \ \ \ \ \ \ \ \ \ \ \ \ \ \ \ \ \ \ \ \ \ \ \ \ \left.+\frac{1}{\lambda^3}e^{-\beta X^+}(1+K)e^{-\beta X^+}(1+K)e^{-\beta X^+}-...\right]\frac{1}{1+K}.\label{eq:pertU}
\end{eqnarray}
It is straightforward to show that $U^{-1}$ takes the form:
\begin{eqnarray}
U^{-1} \lineup = 1+cB(1+K)\left[\frac{1}{\lambda}e^{-\beta X^+} -\frac{1}{\lambda^2}e^{-\beta X^+}K e^{-\beta X^+}\right.\nonumber\\
\lineup\ \ \ \ \ \ \ \ \ \ \ \ \ \ \ \ \ \ \ \ \ \ \ \ \ \ \left.+\frac{1}{\lambda^3}e^{-\beta X^+}Ke^{-\beta X^+}Ke^{-\beta X^+}-...\right]\frac{1}{1+K}.\label{eq:pertUinv}
\end{eqnarray}
Leaving aside the question of convergence of these expansions, it is clear that order-by-order both string fields $U$ and $U^{-1}$ are well-defined and can be used to implement a gauge transformation. Consider a modified gauge parameter $U_N$ given by truncating the expansion of $U$ at the $N$th power of $e^{-\beta x^+}$. The modified gauge parameter does not exactly transform to the tachyon vacuum, but satisfies
\begin{equation}U_N^{-1}(Q+\Psi)U_N = \Psi_\mathrm{tv}+\frac{1}{U_N}\Big[QU_N +\Psi U_N - U_N\Psi_\mathrm{tv}\Big].\label{eq:UNsol}\end{equation}
The gauge transformation removes all powers of $e^{-\beta x^+}$ from $\Psi$ up to and including $e^{-N\beta x^+}$. The remainder term contributes powers $e^{-(N+1)\beta x^+}$ and higher. The order $e^{-(N+1)\beta x^+}$ contribution takes the form
\begin{equation}Q_{\Psi_\mathrm{tv}}\left(cB(1+K)e^{-\beta X^+}\Big[(1+K)e^{-\beta X^+}\Big]^N\frac{1}{1+K}\right).\end{equation}
We know from section \ref{sec:final} that this state behaves for large $N$ roughly as
\begin{equation} N! e^{-(N+1)\beta x^+}.\label{eq:remass}\end{equation}
From this it follows that for any fixed $x^+$, no matter how large, the remainder will diverge in the limit $N\to\infty$, and the gauge transformation fails to map to the tachyon vacuum (or any finite state). This is directly related to the fact that the late time expansion of the solution has zero radius of convergence. Though we can never transform exactly to the tachyon vacuum, we can get ``as close as possible" by choosing $N$ to correspond to the optimal truncation of the late time asymptotic series for a given $x^+$. The optimal truncation is roughly given by
\begin{equation}N_\mathrm{optimal} = e^{\beta x^+}.\end{equation}
Then according to \eq{remass} the remainder will behave as
\begin{equation}e^{- e^{\beta x^+}}.\end{equation}
We cannot conclude based on this result that the remainder represents a physical fluctuation of the tachyon vacuum. All we can conclude is that we cannot get any closer to the tachyon vacuum using a gauge transformation defined perturbatively as an asymptotic series. It is still possible that the gauge transformation to the tachyon vacuum can be defined nonperturbatively.

We have a candidate nonperturbative expression for the gauge parameter $U$ in \eq{Unon}. To transform to the tachyon vacuum, all that is required is to define its inverse. But it turns out that $U$ does not have an inverse. There are several ways to appreciate this fact, but perhaps the most interesting is the existence of an obstruction to invertibility in the form of a nonvanishing projector, called the characteristic projector \cite{EllwoodSing}. The characteristic projector is defined for a related gauge parameter of the form
\begin{equation}V = (1+(\Psi-\Psi_\mathrm{tv})A)(1+A(\Psi-\Psi_\mathrm{tv})).\end{equation}
This can be viewed as a product of two gauge parameters: the first factor is $U$, which transforms $\Psi$ to the tachyon vacuum, and the second factor transforms the tachyon vacuum back into~$\Psi$. Therefore in total $V$ transforms $\Psi$ into itself. But since this transformation ``passes through" the tachyon vacuum, the invertibility of $V$ is directly related to the invertibility of $U$. The characteristic projector of $V$ is defined as
\begin{equation}
X^\infty = \lim_{\eps\to 0}\frac{\eps}{\eps+V}.
\end{equation}
This can be nonzero only if the vanishing factor in the numerator is compensated by a divergence in the denominator; such a divergence should only appear if $V$ fails to have an inverse. Assuming the characteristic projector is contracted with a regular test state, this limit was computed in \cite{singular} for the Kiermaier, Okawa, Soler solution and found to be
\begin{equation}
X^\infty = \sigma\Omega^\infty \overline{\sigma}\frac{1}{1+K},\label{eq:Xinf}
\end{equation}
where $\Omega^\infty$ is the sliver state. We can ask whether this state vanishes. The ghost number $0$ tachyon coefficient in the characteristic projector takes the form:
\begin{equation}
\gamma(x) = \int_0^\infty dt\, e^{-t}\exp\left[-\alpha\,\frac{1+t}{1+2t}\right].
\end{equation}
The integrand is strictly positive, so $\gamma(x)$ does not vanish at any finite time, though it becomes exceedingly small. The late time behavior of the integral can be estimated as
\begin{equation}\gamma(x)\sim \sqrt{\frac{\pi}{2}}\alpha^{1/4} e^{-\alpha/2-\sqrt{\alpha}+1/2}\ \ \ \ (x^+\gg 0).\end{equation}
This implies that it is impossible to completely implement the gauge transformation to the tachyon vacuum at any finite time.

\section{Future Questions}

We conclude by discussing some directions for future investigation.

It would be desireable to extend our analysis to the conventional time-like rolling tachyon deformation. A steepest descent analysis of the coefficients and boundary state could go a long way towards demystifying the late time behavior of this solution. One reason we did not consider this deformation immediately is that the boundary condition changing operators are only known as a perturbative expansion in $e^{X^0}$ \cite{KOS}, and it seems unlikely that the expansion can be resummed into a known function. Still, it should be possible to get insight into the nonperturbative saddle point structure using Pad{\' e} approximants. Another interesting solution is the inhomogeneous rolling tachyon deformation of \cite{Larsen}. Since the associated marginal operator has regular self-OPE, we can describe this background using the analytic solution of Kiermaier, Okawa and Soler. The late time configuration of the string field should describe a periodic array of D-branes; it would be interesting to see how this emerges in string field theory. 

A major theme of our work is understanding the nature of open string deformations of the tachyon vacuum. There are many examples of string field theory solutions which approach the~tachyon vacuum as a function of a parameter. The nature of the implied deformation could be similar to the late time expansion of the light-like rolling tachyon, but this may not always be the case. An example is the $\mathcal{B}_0$ gauge solution for the perturbative vacuum \cite{Schnabl}
\begin{equation}\Psi = \lambda\sqrt{\Omega}c\frac{KB}{1-\lambda\Omega}c\sqrt{\Omega},\ \ \ 0\leq\lambda<1.\end{equation}
Though this solution represents the perturbative vacuum for $|\lambda|<1$, in the limit $\lambda\to 1$ it approaches the tachyon vacuum when expanded into a basis of $L_0$ eigenstates. We can formally expand the solution around the tachyon vacuum:
\begin{equation}
\Psi = \sqrt{\Omega}c\frac{KB}{1-\Omega}c\sqrt{\Omega}\,-\,(1-\lambda)\sqrt{\Omega}c\frac{KB}{(1-\Omega)^2}c\sqrt{\Omega}\,+\,(1-\lambda)^2\sqrt{\Omega}c\frac{KB}{(1-\Omega)^3}c\sqrt{\Omega}\,-\,...\ .
\end{equation}
This expansion is not asymptotic; it is simply undefined, since corrections of order $(1-\lambda)^3$ and higher are divergent.\footnote{Expansion of coefficients in the $L_0$ basis around $\lambda=1$ produces logarithms of $1-\lambda$.} Let us mention a few other examples that may be worth considering: 
\begin{description}
\item{1)} A tachyon lump will smoothly approach the tachyon vacuum far away from its core. Since the boundary state of a lower dimensional D-brane is delta-function localized at a point in the transverse dimension, we might speculate that the expansion of the lump solution around infinity will have finite radius of convergence. This would be interesting to see. 
\item{2)} There is growing evidence \cite{Zwiebach,Maccaferri,MaccaferriMatjej} that a solution representing translation of a reference D-brane orthogonal to its worldvolume will approach the tachyon vacuum as the displacement approaches infinity. If the transverse dimension is noncompact, infinite translation of the D-brane should in some sense leave behind only the tachyon vacuum. This, however, assumes some kind of infrared cutoff whose interpretation is nontrivial from the point of view of open strings attached to the original D-brane. 
\end{description}
In a different direction, we can follow the example of this paper and try to construct nontrivial deformations of the tachyon vacuum in the form of pure gauge asymptotic series.\footnote{An interesting attempt to construct nontrivial fluctuations directly around the tachyon vacuum was discussed in \cite{O-kab}.  The perturbative solution given there has infinite radius of convergence, and so is likely gauge equivalent to the tachyon vacuum.} With some insight, it may be possible to identify the relevant saddle point contributions and use resurgence to resum the trans-series. Various anharmonic oscillations of the tachyon vacuum have been discussed in the context of $p$-adic models \cite{MoellerZwiebach,Hellerman}, and it is possible that similar solutions exist in the full string theory, though their boundary conformal field theory description may be obscure. Perhaps such solutions could even describe perturbative closed string states.

\vspace{.5cm}

\noindent {\bf Acknowledgments} 

\vspace{.25cm}

\noindent We would like to thank the Yukawa Institute for Theoretical Physics at Kyoto University for invitation to  the workshop YITP-T-18-04 ``New Frontiers in String Theory 2018," and specifically for conversations with Masazumi Honda in connection to resurgence theory. This work has been supported in parts by ERDF and M\v{S}MT (Project CoGraDS -CZ.02.1.01/0.0/0.0/15 003/0000437) and by the Czech Science Foundation (GA\v{C}R)
grant 17-22899S.

\begin{appendix}

\section{Matter Correlator}
\label{app:bcc}

In this appendix we compute the matter correlation function \eq{bcc}. First we recall the $(n+1)$-point function of boundary plane wave vertex operators on the upper half plane:
\begin{eqnarray}
\Big\langle e^{ik_0\cdot X}(x_0)...e^{ik_n\cdot X}(x_n)\Big\rangle^{X^0,X^1}_\mathrm{UHP} \lineup = (2\pi)^2\delta^2(k_0+...+k_n+iV)\prod_{i<j}|x_i-x_j|^{2k_i\cdot k_j}\nonumber\\
\lineup\ \ \ \ x_0,...,x_n\in \mathbb{R}.
\end{eqnarray}
The shift in the argument of the delta function is due to the linear dilaton. For string field theory computations it is convenient to transform from the upper half plane to the cylinder $C_L$ of circumference $L$ using the conformal map
\begin{equation}z=\frac{L}{\pi} \tan^{-1}u,\end{equation}
where $u$ is a point on the upper half plane and $z$ is the corresponding point on the cylinder. This gives
\begin{eqnarray}
\Big\langle e^{ik_0\cdot X}(x_0)...e^{ik_n\cdot X}(x_n)\Big\rangle^{X^0,X^1}_{C_L} \lineup = (2\pi)^2\delta^2(k_0+...+k_n+iV)\prod_{i<j}\left|\frac{L}{\pi}\sin\frac{\pi(x_i-x_j)}{L}\right|^{2k_i\cdot k_j}\nonumber\\
\lineup\ \ \ \ x_0,...,x_n\in \mathbb{R}/L.
\end{eqnarray}
We now specialize to the case where all vertex operators except the $0$th take the form $e^{\beta X^+}$. The momenta are then
\begin{equation}
k_{i,\mu} = -i \beta \delta_\mu^+,\ \ \ i=1,...,n.
\end{equation}
Due to the light-like momentum, there are no contractions between pairs of $e^{\beta X^+}$, and the correlator simplifies: 
\begin{eqnarray}
\Big\langle  e^{\beta X^+}(x_1)...e^{\beta X^+}(x_n) e^{ik\cdot X}(x) \Big\rangle^{X^0,X^1}_{C_L}\lineup = (2\pi)^2\delta^2(k-i(n\beta \delta^+ -V))\prod_{i=1}^n\left(\frac{L}{\pi}\sin\frac{\pi(x_i-x)}{L}\right)^{-2}\nonumber\\
\lineup\ \ \ \ x_1,...,x_n,x\in \mathbb{R}/L.\label{eq:appbxpcor}
\end{eqnarray}
With this we can compute 
\begin{equation}
\Tr\Big[\Omega^{\alpha_1}\,\sigma\, \Omega^{\alpha_2}\,\overline{\sigma}\,\Omega^{\alpha_3} e^{-i(k+iV)\cdot X}\Big]^{X^0,X^1}=\Big\langle \sigma(\alpha_2+\alpha_3)\overline{\sigma}(\alpha_3)e^{-i(k+iV)\cdot X}(0)\Big\rangle_{C_L}^{X^0,X^1},
\end{equation}
where $L=\alpha_1+\alpha_2+\alpha_3$. The deformed boundary condition between $\sigma$ and $\overline{\sigma}$ can be represented by repeated integration of the marginal operator along the boundary: 
\begin{equation}
\Tr\Big[\Omega^{\alpha_1}\,\sigma\, \Omega^{\alpha_2}\,\overline{\sigma}\,\Omega^{\alpha_3} e^{-i(k+iV)\cdot X}\Big]^{X^0,X^1}=\left\langle \exp\left[-\lambda\int_{\alpha_3}^{\alpha_2+\alpha_3} dz\, e^{\beta X^+}(z)\right]e^{-i(k+iV)\cdot X}(0)\right\rangle_{C_L}^{X^0,X^1}.
\end{equation}
Expanding the exponential and using \eq{appbxpcor} gives
\begin{eqnarray}
\lineup \Tr\Big[\Omega^{\alpha_1}\,\sigma\, \Omega^{\alpha_2}\,\overline{\sigma}\,\Omega^{\alpha_3} e^{-i(k+iV)\cdot X}\Big]^{X^0,X^1} =\sum_{n=0}^\infty\frac{(-\lambda)^n}{n!}\int_{\alpha_3}^{\alpha_2+\alpha_3} dx_1...\int_{\alpha_3}^{\alpha_2+\alpha_3} dx_n\nonumber\\
\lineup\ \ \ \ \ \ \ \ \ \ \ \ \ \ \ \ \ \ \ \ \ \ \ \ \ \ \ \ \ \ \ \ \ \ \ \ \ \ \ \ \ \ \ \ \ \ \ \ \ \ \ \ \ \ \ \ \ \ \ \ \ \ \ \times\Big\langle e^{\beta X^+}(x_1)...e^{\beta X^+}(x_n) e^{i(k+iV)\cdot X}(0)\Big\rangle^{X^0,X^1}_{C_L}\nonumber\\
\lineup \ \ \ \ = \sum_{n=0}^\infty\frac{(-\lambda)^n}{n!}(2\pi)^2\delta^2(k-i n\beta \delta^+)\left[\int_{\alpha_3}^{\alpha_2+\alpha_3} dx \left(\frac{L}{\pi}\sin\frac{\pi x}{L}\right)^{-2}\right]^n
\nonumber\\
\lineup \ \ \ \ = \sum_{n=0}^\infty\frac{(-\lambda)^n}{n!}(2\pi)^2\delta^2(k-i n\beta \delta^+)\left(\frac{\pi}{L}\frac{\sin\theta_{\alpha_2}}{\sin\theta_{\alpha_1}\sin\theta_{\alpha_3}}\right)^n,\label{eq:appstep}
\end{eqnarray}
where
\begin{equation}\theta_{\alpha_i}\equiv\frac{\pi\alpha_i}{L}.\end{equation}
Next we represent the delta function through the Fourier transform:
\begin{equation}
(2\pi)^2\delta^2(k-i n\beta N^+) = \int d^2 y\, e^{-i k\cdot y} e^{n\beta y^+}.
\end{equation}
Then we can perform the sum in \eq{appstep} underneath the integral over $y$:
\begin{equation}
\Tr\Big[\Omega^{\alpha_1}\,\sigma\, \Omega^{\alpha_2}\,\overline{\sigma}\,\Omega^{\alpha_3} e^{-i(k+iV)\cdot X}\Big]^{X^0,X^1} = \int d^2 y\, e^{-i k\cdot y} \exp\left[-\frac{\pi\lambda}{L}e^{\beta y^+}\frac{\sin\theta_{\alpha_2}}{\sin\theta_{\alpha_1}\sin\theta_{\alpha_3}}\right].
\end{equation}
Taking the Fourier transform gives
\begin{equation}
 \int\frac{d^2 k}{(2\pi)^2} e^{ik\cdot (x- V \ln \frac{\pi}{2})}\Tr\Big[\Omega^{\alpha_1}\,\sigma\, \Omega^{\alpha_2}\,\overline{\sigma}\,\Omega^{\alpha_3} e^{-i(k+iV)\cdot X}\Big]^{X^0,X^1}= \exp\left[-e^{\beta x^+} \frac{2\lambda}{L}\frac{\sin \theta_{\alpha_2}}{\sin\theta_{\alpha_1}\sin\theta_{\alpha_3}}\right],\label{eq:appbcc}
\end{equation}
which is what we wanted to show.

For computing the closed string tadpole amplitude in section \ref{sec:physical}, we need the correlator of boundary condition changing operators with a bulk plane wave vertex operator. Recalling 
\begin{eqnarray}
\lineup\Big\langle e^{i K\cdot X}(z,\overline{z}) e^{ik_1\cdot X}(x_1)...e^{ik_n\cdot X}(x_n)\Big\rangle_\mathrm{UHP}^{X^0,X^1} \nonumber\\
\lineup\ \ \ \ \ \ \ \ \ \ \ \ = (2\pi)^2\delta(K+k_1+...+k_n+iV)|z-\overline{z}|^{\frac{1}{2}K^2}\prod_{i=1}^n|z-x_i|^{2K\cdot k_i}\prod_{i<j}|x_i-x_j|^{2k_i\cdot k_j},\nonumber\\
\lineup \ \ \ \ \ \ \ \ \ \ \ \ \ \ \ \ \mathrm{Im}(z)>0,\ x_1,...,x_n\in \mathbb{R},
\end{eqnarray}
we can follow a similar manipulation as above to show that
\begin{equation}
\int \frac{d^2 k}{(2\pi)^2} e^{ik\cdot x}\Big\langle e^{-i(k+2i V)\cdot X}\left(\frac{i}{2},-\frac{i}{2}\right)\sigma(T)\overline{\sigma}(-T)\Big\rangle_\mathrm{UHP}^{X^0,X^1}=e^{-x^-/\beta}\exp\Big[-4\lambda e^{\beta x^+}\tan^{-1}(2 T)\Big]. 
\end{equation}
This formula is used in the computation of \eq{phys_amp} and \eq{FDelta}.

\section{Coefficients of the Late Time Expansion}
\label{app:f_n}

In this appendix we give a formula for the numbers $f_n$ in \eq{growth} defining the late time expansion of the ghost number zero tachyon. The $f_n$s are determined by the coefficients of the Taylor series expansion of $t(\tau)$ (the inverse function of $\tau(t)$ in \eq{tau}) around the origin $\tau=0$:
\begin{equation}t(\tau) = \sum_{n=1}^\infty t_n \frac{\tau^n}{n!}.\end{equation}
The coefficients $t_n$ are given by
\begin{equation}t_n = \frac{\pi}{2}\frac{1}{n(n+1)}\int_0^\infty d\kappa \, \kappa^{n+1}V_{n-1}^{(1)}(\kappa)e^{-\frac{\pi\kappa}{2}},\label{eq:ln}\end{equation}
where $V_n^{(1)}(\kappa)$ are a sequence of orthogonal polynomials defined by the generating function
\begin{equation}\frac{e^{\kappa \tan^{-1}z}}{1+z^2} = \sum_{n=0}^{\infty} V_n^{(1)}(\kappa)z^n.\end{equation}
They are orthogonal on the interval $(-\infty,\infty)$ with the weight function
\begin{equation}A_1(\kappa)=\frac{\kappa}{\sinh\frac{\pi\kappa}{2}}.\end{equation}
These polynomials are known in the string field theory literature due to their relation with the so-called $\kappa$-basis \cite{spectroscopy}. They are defined by the inner product of $L_0$ and $K_1$ eigenfunctions in the unitary representation $\mathscr{D}_1^+$ of $SL(2,\mathbb{R})$ (a member of the discrete series of unitary representations~\cite{Belov}). In the mathematics literature, they are known as (a modified version of \cite{Mittag-Leffler}) the Mittag-Leffler polynomials \cite{Bateman}. The coefficients \eq{ln} follow from the identity
\begin{equation}\frac{d^2}{d\alpha^2}\alpha\int_{-\infty}^0 d\tau \,t(\tau)e^{-\alpha\tau} = \frac{\pi}{2}\frac{1}{\alpha}\int_0^\infty d\kappa\,\kappa^2\,\frac{ e^{\kappa\tan^{-1}\frac{\kappa}{\alpha}}}{1+(\frac{\kappa}{\alpha})^2}e^{-\frac{\pi\kappa}{2}}\ \ \ \ (\alpha < 0),\end{equation}
which is obtained upon the substitution
\begin{equation}\kappa = \alpha \cot\frac{\pi}{2(t(\tau)+1)}.\end{equation} 
The coefficients $f_n$ are determined by equating
\begin{equation}
\exp\left[-\sum_{n=1}^\infty t_n \frac{\tau^n}{n!}\right]=\sum_{n=0}^\infty f_n \tau^n,
\end{equation}
which implies
\begin{equation}
f_0=1,\ \ \ \ f_n = \frac{1}{n!}\sum_{k=1}^n (-1)^k B_{n,k}(t_1,...,t_{n-k+1}) \ \ \ \ (n\geq1),\label{eq:B8}
\end{equation}
where $B_{n,k}$ are the incomplete exponential Bell polynomials. For computational purposes the following closed form expressions are helpful~\cite{elliptic}:
\begin{eqnarray}
V_{2n}^{(1)}(\kappa) \lineup = \frac{(-1)^n}{(2n)!}\sum_{j=0}^n d_{2j+1}^{(2n+1)}(-1)^j \kappa^{2j},\\
V_{2n+1}^{(1)}(\kappa)\lineup = \frac{(-1)^n}{(2n+1)!}\sum_{j=0}^n d_{2j+2}^{(2n+2)}(-1)^j \kappa^{2j+1},
\end{eqnarray}
where
\begin{equation}d_j^{(n)}=(-1)^n \sum_{k=j}^n \left(\frac{1}{2}\right)^{j-k}L_{n,k}s(k,j),\end{equation}
and where $L_{n,k}$ are Lah numbers and $s(k,j)$ are (signed) Stirling numbers of the first kind. Substituting into \eq{ln} leads to the formulas
\begin{eqnarray}
t_{2n+1}\lineup = \sum_{j=0}^n(-1)^{n+j}d_{2j+1}^{(2n+1)}\frac{(2n+2j+2)!}{(2n+2)!}\left(\frac{2}{\pi}\right)^{2n+2j+2},\\
t_{2n+2}\lineup = \sum_{j=0}^n (-1)^{n+j}d_{2j+2}^{(2n+2)}\frac{(2n+2j+4)!}{(2n+3)!}\left(\frac{2}{\pi}\right)^{2n+2j+4}.
\end{eqnarray}
With this information we can efficiently compute the coefficients $f_n$. The first few values are
\begin{eqnarray}
f_0 \lineup = 1,\nonumber\\
f_1\lineup = -\left(\frac{2}{\pi}\right)^2\approx -.405,\nonumber\\
f_2\lineup = -\frac{1}{2!}\left(\frac{2}{\pi}\right)^43\approx -.246,\nonumber\\
f_3\lineup = -\frac{1}{3!}\left(\frac{2}{\pi}\right)^6\left(19-\frac{\pi^2}{2}\right)\approx -.156,\nonumber\\
f_4\lineup = -\frac{1}{4!}\left(\frac{2}{\pi}\right)^8\Big(191-10\pi^2\Big)\approx -.104,\nonumber\\
f_5\lineup = -\frac{1}{5!}\left(\frac{2}{\pi}\right)^{10}\left(2661-205\pi^2 +\frac{3\pi^4}{2}\right)\approx -.071,\nonumber\\
f_6\lineup = -\frac{1}{6!}\left(\frac{2}{\pi}\right)^{12}\left(47579-4790 \pi^2 +\frac{161\pi^4}{2}\right)\approx -.050,\nonumber\\
f_7\lineup = -\frac{1}{7!}\left(\frac{2}{\pi}\right)^{14}\left(1040047 -\frac{257075 \pi^2}{2} + 3479 \pi^4 + \frac{45 \pi^6}{4}\right)\approx -.035
\end{eqnarray}
Up to a simple prefactor, the coefficients are polynomials in $\pi$ with coefficients which appear to be integers divided by powers of $2$. The requisite powers of $2$ seem to increase slowly with $n$. We observed a cute number-theoretic pattern that the smallest $n$ which requires a factor of $2^p$ in the denominator of a coefficient of the polynomial in $\pi$ is given by
\begin{equation}n=2^{p+1}-1\end{equation}
We have checked this up to $p=4$, confirming that $f_{31}$ is the first time where $16$ appears in a denominator. The coefficients are negative from $n=1$ up to $n=14$, switch to positive from $n=15$ up to $n=87$, and then switch to negative again. The signs likely change an infinite number of times, in accordance with the asymptotic formula \eq{ass_guess}.

\end{appendix}


\begin{thebibliography}{99}

\bibitem{MoellerZwiebach}

  N.~Moeller and B.~Zwiebach,
  ``Dynamics with infinitely many time derivatives and rolling tachyons,''
  JHEP {\bf 0210}, 034 (2002)
  [hep-th/0207107].


\bibitem{Hata}

  M.~Fujita and H.~Hata,
  ``Time dependent solution in cubic string field theory,''
  JHEP {\bf 0305}, 043 (2003)
  [hep-th/0304163].


\bibitem{Schnabl_marg}

  M.~Schnabl,
  ``Comments on marginal deformations in open string field theory,''
  Phys.\ Lett.\ B {\bf 654}, 194 (2007)
  [hep-th/0701248 [HEP-TH]].


\bibitem{KORZ}

  M.~Kiermaier, Y.~Okawa, L.~Rastelli and B.~Zwiebach,
  ``Analytic solutions for marginal deformations in open string field theory,''
  JHEP {\bf 0801}, 028 (2008)
  [hep-th/0701249 [HEP-TH]].

\bibitem{Ellwood}

  I.~Ellwood,
  ``Rolling to the tachyon vacuum in string field theory,''
  JHEP {\bf 0712}, 028 (2007)
  [arXiv:0705.0013 [hep-th]].


\bibitem{KOS}

  M.~Kiermaier, Y.~Okawa and P.~Soler,
  ``Solutions from boundary condition changing operators in open string field theory,''
  JHEP {\bf 1103}, 122 (2011)
  [arXiv:1009.6185 [hep-th]].


\bibitem{Longton}   

	M.~Longton,
  ``Time-Symmetric Rolling Tachyon Profile,''
  JHEP {\bf 1509}, 111 (2015)
  [arXiv:1505.00802 [hep-th]].

\bibitem{SenRolling}

  A.~Sen,
  ``Rolling tachyon,''
  JHEP {\bf 0204}, 048 (2002)
  [hep-th/0203211].


\bibitem{Hellerman}

  S.~Hellerman and M.~Schnabl,
  ``Light-like tachyon condensation in Open String Field Theory,''
  JHEP {\bf 1304}, 005 (2013)
  [arXiv:0803.1184 [hep-th]].

\bibitem{Hellerman1}

  S.~Hellerman and I.~Swanson,
  ``Cosmological solutions of supercritical string theory,''
  Phys.\ Rev.\ D {\bf 77}, 126011 (2008)
  [hep-th/0611317].

\bibitem{Hellerman2}

  S.~Hellerman and I.~Swanson,
  ``Dimension-changing exact solutions of string theory,''
  JHEP {\bf 0709}, 096 (2007)
  [hep-th/0612051].


\bibitem{Hellerman3}

  S.~Hellerman and I.~Swanson,
  ``Cosmological unification of string theories,''
  JHEP {\bf 0807}, 022 (2008)
  [hep-th/0612116].



\bibitem{SenTM}

  A.~Sen,
  ``Tachyon matter,''
  JHEP {\bf 0207}, 065 (2002)
  [hep-th/0203265].
  
\bibitem{Okawa}

  Y.~Okawa,
  ``Comments on Schnabl's analytic solution for tachyon condensation in Witten's open string field theory,''
  JHEP {\bf 0604}, 055 (2006)
  [hep-th/0603159].

\bibitem{simple}

  T.~Erler and M.~Schnabl,
  ``A Simple Analytic Solution for Tachyon Condensation,''
  JHEP {\bf 0910}, 066 (2009)
  [arXiv:0906.0979 [hep-th]].


\bibitem{Schnabl}

  M.~Schnabl,
  ``Analytic solution for tachyon condensation in open string field theory,''
  Adv.\ Theor.\ Math.\ Phys.\  {\bf 10}, no. 4, 433 (2006)
  [hep-th/0511286].



\bibitem{Moeller}

  F.~Beaujean and N.~Moeller,
  ``Delays in Open String Field Theory,''
  arXiv:0912.1232 [hep-th].


\bibitem{KOSsing}

  T.~Erler and C.~Maccaferri,
  ``String Field Theory Solution for Any Open String Background,''
  JHEP {\bf 1410}, 029 (2014)
  [arXiv:1406.3021 [hep-th]].

\bibitem{IdSing}

  T.~Erler,
  ``The Identity String Field and the Sliver Frame Level Expansion,''
  JHEP {\bf 1211}, 150 (2012)
  [arXiv:1208.6287 [hep-th]].

\bibitem{Miller} P. Miller, {\it Applied Asymptotic Analysis}, American Mathematical Society, Providence, 2006.

\bibitem{Kawano}

  T.~Kawano, I.~Kishimoto and T.~Takahashi,
  ``Gauge Invariant Overlaps for Classical Solutions in Open String Field Theory,''
  Nucl.\ Phys.\ B {\bf 803}, 135 (2008)
  [arXiv:0804.1541 [hep-th]].

\bibitem{boundarystate}   M.~Kudrna, C.~Maccaferri and M.~Schnabl,
  ``Boundary State from Ellwood Invariants,''
  JHEP {\bf 1307}, 033 (2013)
  [arXiv:1207.4785 [hep-th]].

\bibitem{tadpole}

  I.~Ellwood,
  ``The Closed string tadpole in open string field theory,''
  JHEP {\bf 0808}, 063 (2008)
  [arXiv:0804.1131 [hep-th]].


\bibitem{Shatashvilli}

  A.~A.~Gerasimov and S.~L.~Shatashvili,
  ``On exact tachyon potential in open string field theory,''
  JHEP {\bf 0010}, 034 (2000)
  [hep-th/0009103].


\bibitem{Moore}

  D.~Kutasov, M.~Marino and G.~W.~Moore,
  ``Some exact results on tachyon condensation in string field theory,''
  JHEP {\bf 0010}, 045 (2000)
  [hep-th/0009148].

\bibitem{Ecalle} J. {\'E}calle, {\it Les fonctions resurgentes}, I, II et III, Publications Math{\'e}matiques d'Orsay, 1981 et 1985.

\bibitem{Unsal}

  A.~Cherman, D.~Dorigoni and M.~Unsal,
  ``Decoding perturbation theory using resurgence: Stokes phenomena, new saddle points and Lefschetz thimbles,''
  JHEP {\bf 1510}, 056 (2015)
  [arXiv:1403.1277 [hep-th]].


\bibitem{Marino}

  M.~Mari{\~n}o,
  ``Lectures on non-perturbative effects in large $N$ gauge theories, matrix models and strings,''
  Fortsch.\ Phys.\  {\bf 62}, 455 (2014)
  [arXiv:1206.6272 [hep-th]].

\bibitem{Schiappa}   
	
  I.~Aniceto, G.~Basar and R.~Schiappa,
  ``A Primer on Resurgent Transseries and Their Asymptotics,''
  arXiv:1802.10441 [hep-th].

\bibitem{EllwoodSing}

 I.~Ellwood,
  ``Singular gauge transformations in string field theory,''
  JHEP {\bf 0905}, 037 (2009)
  [arXiv:0903.0390 [hep-th]].

\bibitem{singular}

  T.~Erler and C.~Maccaferri,
  ``Connecting Solutions in Open String Field Theory with Singular Gauge Transformations,''
  JHEP {\bf 1204}, 107 (2012)
  [arXiv:1201.5119 [hep-th]].

\bibitem{Larsen}

  F.~Larsen, A.~Naqvi and S.~Terashima,
  ``Rolling tachyons and decaying branes,''
  JHEP {\bf 0302}, 039 (2003)
  [hep-th/0212248].

\bibitem{Zwiebach}

  B.~Zwiebach,
  ``A Solvable toy model for tachyon condensation in string field theory,''
  JHEP {\bf 0009}, 028 (2000)
  [hep-th/0008227].


\bibitem{Maccaferri}   
  C.~Maccaferri and M.~Schnabl,
  ``Large BCFT moduli in open string field theory,''
  JHEP {\bf 1508}, 149 (2015)
  [arXiv:1506.03723 [hep-th]].
  
\bibitem{MaccaferriMatjej}

  M.~Kudrna and C.~Maccaferri,
  ``BCFT moduli space in level truncation,''
  JHEP {\bf 1604}, 057 (2016)
  [arXiv:1601.04046 [hep-th]].
  
\bibitem{O-kab}

  O.~K.~Kwon,
  ``Marginally Deformed Rolling Tachyon around the Tachyon Vacuum in Open String Field Theory,''
  Nucl.\ Phys.\ B {\bf 804}, 1 (2008)
  [arXiv:0801.0573 [hep-th]].

\bibitem{spectroscopy} 

  L.~Rastelli, A.~Sen and B.~Zwiebach,
  ``Star algebra spectroscopy,''
  JHEP {\bf 0203}, 029 (2002)
  [hep-th/0111281].

\bibitem{Belov}

  D.~M.~Belov and C.~Lovelace,
  ``Witten's vertex made simple,''
  Phys.\ Rev.\ D {\bf 68}, 066003 (2003)
  [hep-th/0304158].


\bibitem{Mittag-Leffler} 

M.~S.~Stankovi{\'c}, S.~D.~Marinkovi{\'c}, P.~M.~Rajkovi{\'c}, 
``The deformed and modified Mittag-Leffler polynomials," 
Math. Comput. Model. {\bf 54} (2011) 721-728.

\bibitem{Bateman} 

	H.~Bateman,
	``The polynomial of Mittag-Leffler,''
	Proc. Natl. Acad. Sci. {\bf 26} (1940) 491-496
	
	
\bibitem{elliptic}
	
	J.~S.~Lomont, {\it Elliptic Polynomials.} Boca Raton, FL: Chapman \& Hall/CRC, 2001. Chapter~11.


\end{thebibliography}
\end{document}